\documentclass[12pt, a4paper]{article}
\usepackage[utf8]{inputenc}

\usepackage[margin=1in]{geometry}
\usepackage{mathtools}
\usepackage{amsmath}
\usepackage{amsthm}
\usepackage{amssymb}
\usepackage{setspace}
\usepackage{booktabs}
\usepackage{makecell}
\usepackage{tabularx}
\usepackage{amsmath} 
\usepackage{amssymb}
\usepackage{bm}
\usepackage{ascmac}
\usepackage{setspace}
\usepackage[at]{easylist}
\usepackage[normalem]{ulem}
\usepackage{comment}

\mathtoolsset{showonlyrefs}  

\makeatletter

\usepackage{amsthm}
\usepackage{amsfonts}
\usepackage{bbm}
\usepackage{graphics}
\usepackage{graphicx}
\usepackage{enumerate}
\usepackage{float}
\usepackage{caption}
\usepackage{subcaption}
\usepackage{natbib}
\usepackage[dvipsnames]{xcolor}

\theoremstyle{plain}

\theoremstyle{remark}

\newtheorem{remark}{Remark}

\usepackage{booktabs}        
\usepackage{multirow}        
\usepackage{graphicx}        
\usepackage{subcaption}      
\usepackage{threeparttable}  
\usepackage{bm}              
\usepackage{enumitem}        
\usepackage{xspace}

\newcommand{\alg}[1]{\textbf{\textsc{#1}}\xspace}
\newcommand{\Greedy}{\alg{Greedy}}
\newcommand{\SQ}{\alg{SQ}}
\newcommand{\AQ}{\alg{AQ}}
\newcommand{\TEC}{\alg{TEC}}

\usepackage[hyphens]{url}
\usepackage{hyperref}
\hypersetup{
  colorlinks=true,
  linkcolor=BrickRed,
  citecolor=BrickRed,
  filecolor=BrickRed,
  urlcolor=BrickRed,
  hypertexnames=true
}

\theoremstyle{plain}

\usepackage{algorithm,algpseudocode}

\usepackage{amsmath}
\DeclareMathOperator*{\argmax}{arg\,max}

\allowdisplaybreaks

\makeatother

\title{Designing Recommendation Exposure\\ and Favorite Lists:\\ A Field Experiment in a Spot-Work Platform\footnote{We are grateful to Yu Awaya, Tomohiro Hara, Keisuke Horikoshi, Kenzo Imamura, Michihiro Kandori, Daiji Kawaguchi, Fuhito Kojima, Takahiro Moriya, Fumio Ohtake, Shusaku Sasaki, Uta Sch{\"o}nberg, and all participants of the ERATO Kojima Market Design Project online seminar for helpful comments. This work has been supported by JST ERATO JPMJER2301, JST PRESTO JPMJPR2368, and JSPS KAKENHI 25K16620, Japan, and Timee, Inc. We are grateful to Kazuki Takaishi and Takahide Kimura for sharing the data and for their technical and institutional knowledge of the Timee platform. The study protocol was approved by the Institutional Review Board of the University of Tokyo (IRB No.~E25ALS0255) and pre-registered in the AEA RCT Registry (AEARCTR-0017620). The views expressed in this paper are those of the authors and do not necessarily reflect the views of Timee.}}
\author{
  Kazuki Sekiya\thanks{\href{mailto:kazuki-sekiya@g.ecc.u-tokyo.ac.jp}{kazuki-sekiya@g.ecc.u-tokyo.ac.jp}, Graduate School of Economics, The University of Tokyo} \and
  Suguru Otani\thanks{\href{mailto:suguru.otani@e.u-tokyo.ac.jp}{suguru.otani@e.u-tokyo.ac.jp}, Graduate School of Economics, The University of Tokyo} \and
  Yuki Komatsu\thanks{\href{mailto:komatsu-yuki286@g.ecc.u-tokyo.ac.jp}{komatsu-yuki286@g.ecc.u-tokyo.ac.jp}, Department of Economics, The University of Tokyo} \and
  Yuki Fujii\thanks{\href{mailto:yuki\_fujii@timee.co.jp}{yuki\_fujii@timee.co.jp}, Timee, Inc., Tokyo, Japan} \and
  Shunsuke Ozeki\thanks{\href{mailto:shunsuke.ozeki@timee.co.jp}{shunsuke.ozeki@timee.co.jp}, Timee, Inc., Tokyo, Japan} \and
  Shunya Noda\thanks{Contact Author. \href{mailto:shunya.noda@e.u-tokyo.ac.jp}{shunya.noda@e.u-tokyo.ac.jp}, Graduate School of Economics, The University of Tokyo}
}

\date{\today}

\begin{document}

\singlespacing

\maketitle

\begin{abstract}
How should recommender systems be designed when recommendations shape access to scarce, short-lived opportunities? We study this question in a production setting: \emph{Timee}, Japan's largest platform for spot work, where workers favorite job templates and receive notifications when firms post shifts from those templates. Maximizing predicted favoriting can generate misdirected concentration: recommendations accumulate on popular templates that create few viable job openings, while templates with unmet labor demand receive too little exposure. We design exposure-control mechanisms for favorite-list management, reallocating template exposure based on posting activity and unfilled capacity. The proposed recommender, \emph{thresholded eligibility control} (\TEC), is fully parallelizable and suitable for large-scale digital platforms. In simulations calibrated to Timee data, \TEC raises the per-round job-finding rate from 57.6\% to 70.0\%. A prefecture-level randomized field experiment increases realized matches and exposure per active template, reduces the share of low-exposure templates, and improves impression-level favoriting and downstream matching.
\end{abstract}

\onehalfspacing

\section{Introduction}

Recommender systems have become a central feature of digital platforms, helping users navigate large sets of options. By predicting users' preferences and highlighting promising alternatives, they reduce information frictions and facilitate better matches between users and options. At the same time, the choice of which candidates to recommend---namely, how \emph{exposure} is allocated across options---directly affects which matches are ultimately realized. When viewed through this lens, recommender system design is fundamentally about market design for allocating limited exposure, which can significantly shape matching outcomes.

In recent years, online labor-matching platforms have rapidly expanded worldwide by offering ``spot work''---short-hour, one-off gigs facilitated digitally---spanning services such as Taskmo (India), TROOPERS (Singapore), Geupgu (South Korea), QuikShift (Australia), Coople (Switzerland/UK), and Instawork (United States), in line with a broader expansion into flexible, on-demand labor markets \citep{kassi2018online,kanayama2024nonparametric}.\footnote{This definition follows that of the Japan Spot Work Association.} As of 2024, Japan's leading spot-work platform, \emph{Timee}, supports over one million active users and vacancies per month across industries such as food service, retail, and logistics. By enabling workers to quickly access short-term jobs and allowing firms to flexibly fill temporary labor needs, these platforms can reduce frictions in local labor markets and improve the utilization of available labor capacity.

In spot-work markets, the role of the recommender system and exposure allocation is crucial for the matching outcome. Workers typically access the platform briefly and intermittently, face a large number of largely homogeneous job postings, and have weak incentives to evaluate individual opportunities carefully. Moreover, spot-work contracts do not presuppose long-term relationships between workers and firms, which limits the extent to which workers can rely on relationship-based learning or repeated interactions when making job choices. As a result, workers rely heavily on the information provided by the platform when deciding which jobs to apply for. In addition, job postings in spot-work markets are short-lived and often filled quickly, which makes the timely presentation of relevant opportunities essential for successful matching. Taken together, recommender design is central to workers' and firms' welfare and the platform's overall performance.

In this paper, we study how the platform should allocate exposure on spot-work platforms to improve market-level matching outcomes. In contrast to traditional objectives that focus on optimizing conversion metrics such as click-through rates, we take the realized \emph{per-round job-finding rate}, the share of active job-seeking workers who successfully match with an offering for that date, as the primary outcome of interest. To study this question, we develop a model that closely mirrors the matching process implemented in practice on the Timee platform. Using this framework, we demonstrate that a commonly used recommender that myopically optimizes conversion metrics can induce excessive and misdirected concentration of exposure, undermining matching outcomes. We then propose a novel and practical recommender design to address this issue and evaluate its performance through simulations calibrated to detailed operational data from Timee, as well as through a large-scale field experiment conducted on the platform.

This paper makes four main contributions. First, we formulate a model of the spot-work market that explicitly captures the process through which workers receive recommendations for job templates, form favorite lists, and apply to offerings, allowing us to directly study how recommender system design affects matching outcomes. We show that myopic optimization of favoriting probabilities often causes overly concentrated exposure allocation. Moreover, we demonstrate that in this environment, the problem is not only the degree of concentration but also its direction: exposure tends to concentrate on job templates that are favored by workers yet correspond to relatively few posted offerings. As a result, optimizing favoriting probabilities alone does not necessarily ensure the fulfillment of the primary objective in the spot-work market: providing workers with the actual opportunity to obtain employment that aligns with their preferences, abilities, and availability. Although our empirical setting is a spot-work platform, the underlying design problem is more general: when recommendations shape access to scarce or time-sensitive opportunities, optimizing intermediate engagement metrics can be misaligned with the platform's ultimate matching objective.

Second, we show that introducing exposure quotas is an effective way to mitigate inefficient concentration in spot-work platforms, and we propose a method for adjusting these quotas adaptively over time. Rather than fixing quotas ex ante, our approach updates per-template exposure limits based on recent labor demand and success in filling posted offerings, allowing the recommender system to respond to changing market conditions. Moreover, we explicitly design the proposed method to be fully parallelizable, reflecting the scalability requirements of production recommender systems. This feature is essential for a large platform such as Timee, which operates at the scale of over one million active users and vacancies per month, and is similarly important for other large-scale platforms that allocate recommendations in real time.

Third, we evaluate the proposed recommenders through simulations calibrated to detailed operational data from the Timee platform. A key purpose of this exercise is to assess, before live deployment, which recommender designs are likely to improve market-level outcomes and are suitable for implementation on the platform. Using observed distributions of job postings, favoriting behavior, and worker application patterns, we construct a simulation environment that closely matches the qualitative characteristics of a real spot-work market. Within this setting, a conventional recommender that optimizes favoriting behavior (\Greedy) achieves a per-round job-finding rate of 57.6\%, while introducing static quotas on exposure (\SQ) improves this rate to 64.2\%. The adaptive quota method (\AQ) and our deployable parallelized implementation (\TEC, thresholded eligibility control) raise the per-round job-finding rate further to 69.5\% and 70.0\%, respectively. These gains arise from reallocating wasted exposure away from templates with limited available offerings and toward templates with substantial unmet labor demand. The proposed adaptive approach also consistently outperforms the benchmark policies across a wide range of parameter values and market conditions, indicating that the improvements reflect a robust enhancement in market-level matching efficiency rather than fine-tuned parameter choices. These results helped identify \TEC as the most promising design for the subsequent production-scale rollout.

Fourth, we deploy the proposed recommender (\TEC) in a live field experiment and evaluate its causal effects. To address user-level interference and cross-unit spillovers, which are well-known problems in matching platform experimentation \citep{manshadi2023redesigning}, we adopt a prefecture-level randomized rollout analyzed with difference-in-differences (DID) style regressions: Aomori serves as the treatment region and Iwate as a geographically proximate but separate labor-market control, with the algorithm introduced as the default recommender for Aomori only. The intervention window spans January 12--February 11, 2026, and we examine post-period action outcomes such as matches via recommendations during the intervention.

For the prefecture-by-day analysis, outcomes include both market-level outcomes, such as daily favorites and matches, and simulation-analogue outcomes---average recommended workers per template, subscribers per template, and the active-template \emph{per-round fill rate}, matched slots over posted capacity among active templates---and we estimate intention-to-treat (ITT) effects with prefecture and date fixed effects. The rollout increased realized matches and recommendation exposure per template, while leaving daily favorites largely unchanged. A distribution-regression DID further shows that \TEC reduced the share of active template-days with very low recommendation counts, indicating that the algorithm shifted exposure away from the lower tail of the template-level distribution. Moreover, the distributional estimate shows a reduction in the zero-subscriber mass of the template-day distribution. These findings support the simulation's central prediction that reallocating exposure toward templates with unmet labor demand can improve market-level matching outcomes. The absence of detectable movements in average subscribers per template and active-offering fulfillment is also consistent with the mechanism: these outcomes depend on the gradual accumulation of favorite-list stocks and therefore need not respond fully within a one-month rollout. We therefore use transition simulations to study the adjustment path beyond the experimental window, and find that longer implementation plausibly allows the exposure reallocation to feed through to subscriber stocks and downstream fulfillment.

In user-impression analyses, we model a worker's decision to (i) favorite a recommended template and (ii) match with an offering from a favorited template as binary discrete choice problems, and estimate treatment-control contrasts for these behaviors. The results show that \TEC improves the effectiveness of individual recommendation exposures: treated impressions are more likely to generate favorites and, conditional on favoriting, more likely to lead to matches. The improvement in match rates is consistent with the mechanism that \TEC steers exposure toward offerings with greater unmet labor demand, thereby increasing the likelihood of successful matching. By contrast, the improvement in the impression-to-favorite margin is less expected. A natural hypothesis is that workers are more likely to favorite templates with more open offerings, but controlling for the presence of open offerings does not attenuate the estimated favoriting effect. Instead, the evidence points to recommendation diversity as the primary driver: \TEC reduces repeated exposure to the same template, and the treatment effect on favoriting increases substantially once repeat viewing is controlled for. These impression-level findings provide an empirical perspective that is not captured in the simulation and suggest that reducing misdirected concentration enhances not only aggregate exposure allocation but also the quality and novelty of recommendation lists experienced by individual users.


Although our analysis mainly focuses on spot-work platforms, the implications of our findings extend well beyond this specific setting. A central insight of the paper is that recommender systems in markets with short-lived opportunities and capacity constraints should be viewed as allocation devices that shape realized matches, rather than merely as tools for reducing information friction. In such environments, optimizing intermediate conversion metrics, such as clicks or favorites, can be misaligned with final market outcomes and may lead to congestion and wasted supply. Our approach for allocating exposure illustrates a general design principle for addressing this misalignment by incorporating supply-side information and adjusting recommendations in a state-dependent manner. This perspective is relevant for a wide range of platforms, including online dating, tutoring and mentoring marketplaces, and appointment scheduling systems, where timely matching and limited capacity play a central role. More broadly, the paper highlights how combining market-design objectives with scalable recommender system implementations can improve welfare in digitally mediated matching markets.


\paragraph{Related Literature}

Research on how Internet-based job search affects labor market outcomes has expanded steadily since the early 2000s \citep{autor2001wiring,kuhn2004internet}. The review by \citet{kuhn2014ineffective} concludes that search-centered approaches that merely provide information, such as online job boards, have limited effects on employment outcomes. More recent work, however, emphasizes that platforms can improve labor market performance by actively guiding search through recommendations. In particular, studies in online labor markets show that well-designed recommender systems can improve hiring, match formation, or search efficiency \citep{horton2017effects,reusens2018evaluating,belot2019providing}. The role of recommendation is particularly important in spot-work platforms, where matches are extremely short-lived and formed at high frequency, making the allocation of exposure a central design problem. Moreover, macro indicators in the Japanese context suggest that, in recent years, hires facilitated through spot-work platforms have outpaced part-time placements via public employment offices \citep{kanayama2024nonparametric}, while related evidence documents the rise of full-time hiring intermediated by digital platforms \citep{otani2025onthejob}. This shift underscores the growing policy and economic relevance of platform design for short-duration jobs, making Timee's comprehensive, high-frequency data an ideal setting to study how recommendation and exposure allocation translate into market-level matching outcomes.

A growing literature also highlights that recommender systems optimized for short-term engagement metrics, such as clicks or views, may fail to improve and can even worsen market-level outcomes in labor markets. When a recommendation helps one worker obtain a job by crowding out others competing for the same position, aggregate employment need not increase. The empirical relevance of such displacement and congestion effects has been documented not only in standard online labor-market settings with job seekers \citep{horton2024reducing} but also in job search assistance and recommendation contexts \citep{crepon2013labor,gautier2018estimating,behaghel2024potential}. Motivated by these concerns, recent studies propose recommender designs that explicitly limit the concentration of exposure on a small set of jobs, thereby mitigating congestion and improving efficiency and fairness at the market level \citep{naya2021designing,manshadi2023redesigning}. Related ideas also appear in dating markets, which share the feature that both sides' preferences matter for match formation. In that literature, several papers design recommendations based on stable matchings derived from transferable-utility models \citep{tomita2022matching,tomita2023fast,chen2023reducing}. Our paper shares the goal of avoiding excessive concentration of recommendations on popular templates to improve overall market efficiency. However, our study differs from these studies in two key respects: first, we focus on a spot-work market where matches are instantly formed in large volumes at high frequency, which requires highly parallelizable and computationally efficient implementations; second, we propose an adaptive method that adjusts recommendation parameters to heterogeneous local market conditions.

Recent work has utilized field experiments to study algorithmic recommendation and ranking policies in matching markets, using designs that range from individual-level A/B tests to clustered implementations to address equilibrium effects. Early platform work studies how algorithmic recommendations shape search and matching in online labor markets \citep{horton2017effects}, while more recent experiments in two-sided markets such as online dating and volunteer platforms explicitly redesign exposure/ranking rules and often rely on region-level variation to mitigate spillover concerns \citep{rios2023improving,manshadi2023redesigning,sekiya2026integrating}. In the policy domain, a contemporaneous wave of projects partners with public employment services (PES) to embed digital job-search assistance for unemployed workers into administrative platforms \citep{behaghel2024potential,altmann2022direct,belot20222longterm,dhia2022can,le2023can}.

Relative to these prior works, our experiments have two distinctive features. First, we study a spot-work market in which matches occur at high frequency and at a large scale. This contrasts with PES-based experiments that predominantly focus on long-term employment and unemployed workers. While long-term contracts still dominate labor markets, the share of spot work is growing, making this setting economically relevant in its own right. Moreover, from the perspective of recommender system experiments, high-volume and high-frequency matching environments are common---not only in gig work but also in e-commerce, online dating, and related platforms---thus our findings have implications that extend naturally to a broader class of markets. Second, thanks to the full cooperation of a nationwide platform, we implement a production-grade recommender as the default policy at the prefecture level and estimate ITT effects using a randomized rollout analyzed with DID-style regressions. This prefecture-level rollout directly addresses interference in a shared marketplace and lets us evaluate the same market-level allocation objects emphasized in the simulation. Furthermore, the rollout provides impression-level evidence on live-platform user responses. The impression-level data allow us to observe which templates were actually shown to each worker, thereby helping mitigate a common ``choice-set problem'' in classical job-search data: researchers often observe final choices, but not workers' choice sets.\footnote{Several recent studies \citep[e.g.,][]{stanton2021benefits,roussille2025bidding} also exploit detailed platform data to study job-search behavior and labor-market outcomes.}


Our paper also contributes to the literature on digital conversion and platform design by studying a mechanism that has been largely overlooked in both e-commerce and matching markets: \emph{favorite-list management} as a dynamic conversion device for capacity-constrained goods. Various preceding works have shown that digital platforms can raise purchases by improving how users return to and act on high-intent options, through rankings and search design \citep{ursu2018power,compiani2024online}, clickstream-based prediction of downstream purchase behavior \citep{montgomery2004modeling}, cart-based targeting of short-listed products \citep{luo2019when}, interface designs that induce choice closure at the postdecision-prepurchase stage \citep{lee2025push}, and checkout simplification \citep{unal2023fewer}. This literature shows that platform design affects conversion not only by better predicting preferences, but also by managing intermediate states of user intent. However, these studies largely examine retail environments in which the object of interest remains purchasable. By contrast, we study a setting in which users save objects precisely because capacity is intermittent and perishable, so conversion depends on future supply arrivals and competition from other users. This makes favorite-list management a problem of dynamic allocation under scarcity. We show that the favorite list should be understood as a stock of latent demand that mediates the relationship between platform exposure and realized transactions, making it a cumulative channel through which recommendations can affect long-run match formation. To our knowledge, this is the first paper to study the conversion effect of favorite-list management for capacity-constrained goods.

\section{Institutional Details}

This section describes the institutional features of Timee that are central to our recommender design and evaluation. Although our approach is relevant to a broader class of capacity-constrained, short-lived matching environments (e.g., task gigs, appointment slots, reservations), its implementation depends on how the platform converts exposure into matches. Therefore, we summarize Timee's platform features before turning to the simulation model, algorithm design, and field evaluation.

\subsection{About Timee}
Timee is a Japanese spot-work platform where workers can register for free and browse and apply to short-term offerings without a traditional hiring process (e.g., no resume screening as a prerequisite for applying). As of 2024, more than 7 million workers were registered on the platform, exceeding 10\% of Japan's labor force of roughly 70 million, illustrating its significant penetration in the country's labor market \citep{kanayama2024nonparametric}.
Timee connects businesses posting short-term, on-site jobs with workers seeking flexible, on-demand work opportunities through its app.

On the labor-demand side, firms register after verification and post shifts (tasks, wages, times, and workplace) via a simple interface; listings become immediately visible, and matching and contract procedures are completed automatically on the platform.
On the labor-supply side, workers register for free and can apply across sectors (food, retail, logistics) without traditional hiring procedures---no formal applications, interviews, or long-term contracts. Although some postings require certified qualifications (e.g., nurses, caregivers, childcare), these are a relatively small share; most jobs indicate minimal skill requirements and are broadly accessible. The platform monetizes only completed matches, collecting a 30\% commission on the contracted salary-based payment from businesses, which aligns incentives toward increasing successful, timely fulfillment rather than paid access or advertising. This arrangement benefits both companies and workers by offering flexible, short-term employment opportunities while avoiding the formalities and commitments of long-term contracts.

\subsection{Offering}

At Timee, workers and firms are matched at the \emph{offering} level, where an offering is a vacancy posting for a specific shift. Each offering represents a shift (lasting from one hour to a full day) at a specific work date and location with the wage and other amenities specified as in a standard job posting. Workers may repeatedly work on offerings posted by the same firm, but they are under no obligation to do so. Likewise, firms are under no obligation to post similar offerings regularly. When a worker applies to an offering, the match is formed immediately, without waiting for any further response from the firm. Accordingly, firms do not screen or select among applicants and accept workers on a first-come, first-served (FCFS) basis.
\footnote{While firms do not screen applicants, the platform has mechanisms to address repeated late cancellations and no-shows that could undermine the reliability of confirmed shifts.}
The FCFS rule is necessary in spot-work markets, where job postings and applications occur over very short time horizons, to ensure that workers can finalize their schedules immediately.\footnote{For detailed offering-level summary statistics on Timee, see \citet{kanayama2026justminimumwagehikes}, who analyze the short-run effects of Japan's minimum wage revision on the Timee platform. Note that, although the revision occurred near our study period, our field implementation began about three months after the policy change, which mitigates concerns about the immediate adjustment effects.}

\subsection{Template and Favorite List}

\begin{figure}[!tp]
    \begin{center}
        \begin{subfigure}{0.48\textwidth}
            \centering
            \includegraphics[width=\textwidth]{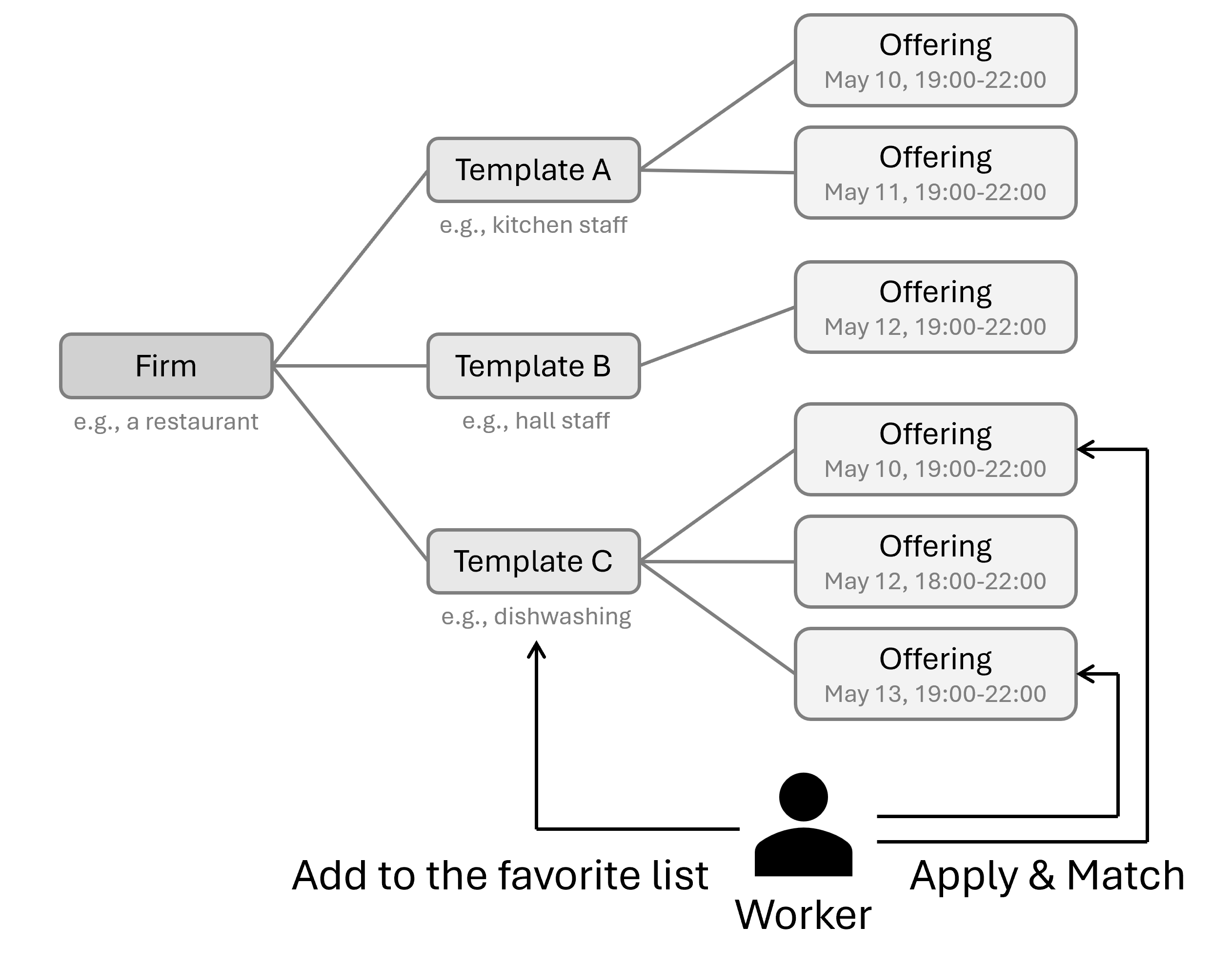}
            \caption{Relationship among Firms, Templates, Offerings, and Workers}
            \label{fig:firm_template_offering}
        \end{subfigure}
        \hfill
        \begin{subfigure}{0.48\textwidth}
            \centering
            \includegraphics[width=\textwidth]{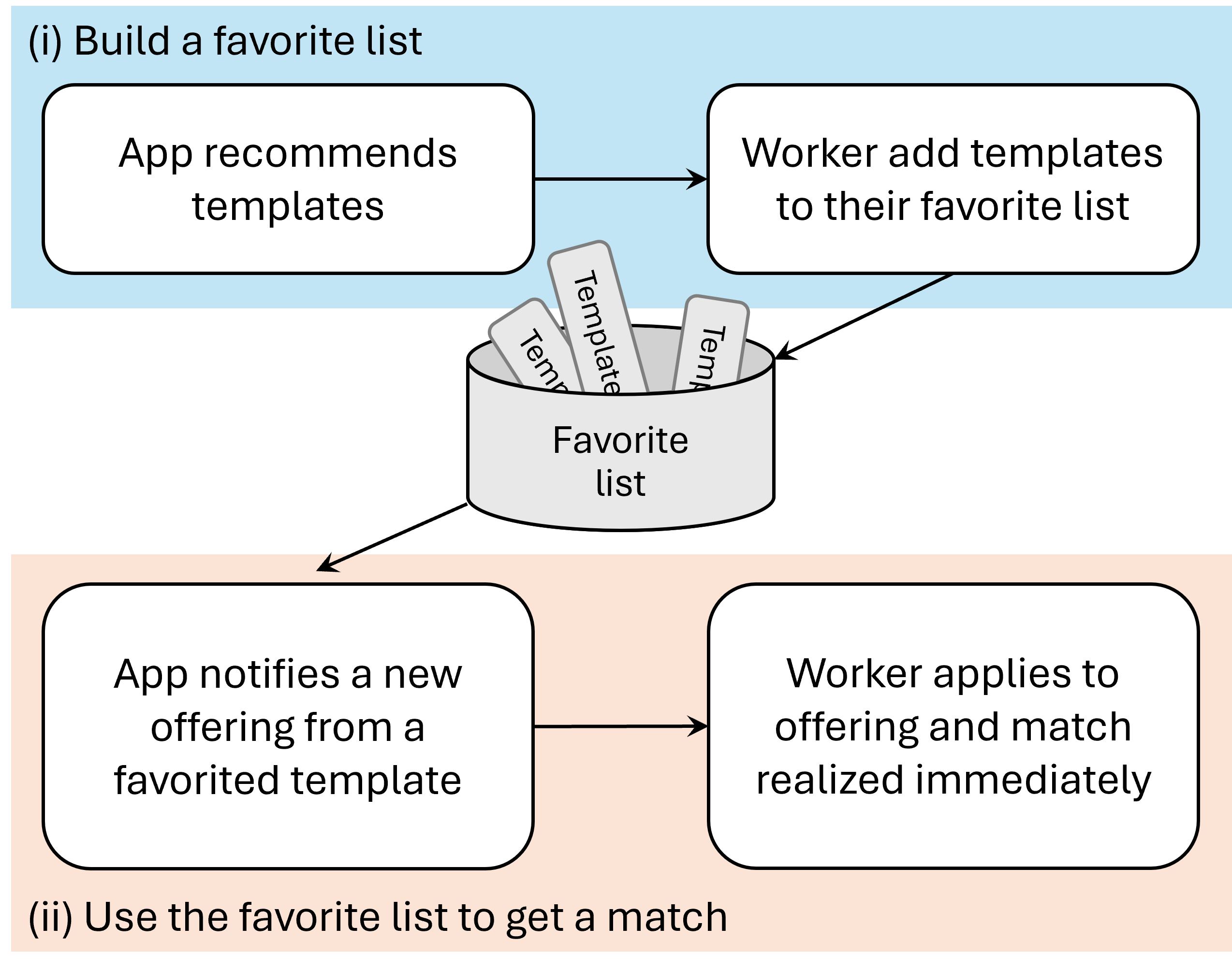}
            \caption{How Workers Build and Use Favorite Lists to Match with Offerings}
            \label{fig:flow_from_recommendation_to_match}
        \end{subfigure}
    \end{center}
    \caption{Templates, Offerings, and the Path from Recommendations to Matches}
    \label{fig:flow_from_search_to_match}
\end{figure}

Firms typically post large numbers of similar offerings, often across different days. On Timee, rather than creating each offering from scratch, firms first create a \emph{template} that describes the underlying job, and then post individual offerings---corresponding to specific dates and time slots---based on the template. It is common for firms to maintain multiple templates representing different job types. As illustrated in Figure~\ref{fig:firm_template_offering}, for example, a restaurant may prepare separate templates for kitchen staff, hall staff, and dishwashing.

Because firms may repeatedly generate offerings from the same templates, Timee encourages workers to construct a \emph{favorite list} consisting of templates they are interested in, so that they can more easily access future offerings derived from those templates. For ease of exposition, we refer to workers who have favorited a given template as the \emph{subscribers} of that template. Subscribers receive notifications when new offerings are generated from the favorited template.

On Timee, offerings posted by popular firms tend to be filled very rapidly, which makes it difficult for workers who search for jobs only intermittently to secure a match through browsing alone. Consequently, constructing a favorite list plays a critical role in enabling workers to access employment opportunities on the platform. Thus, improving matching outcomes in this setting requires more than the ranking, search-design, and interface interventions emphasized in prior work. It also requires managing workers' favorite lists so that the stock of latent worker demand is aligned with the future offering capacity of templates.

\subsection{Template Recommendation}

To assist workers in building their favorite lists, Timee provides a personalized recommender system that suggests templates likely to be of interest to each worker. Although workers may find offerings and add templates to their favorite lists through several channels, this paper focuses on Timee's template recommendation function, which suggests templates for workers to favorite.\footnote{For example, workers may directly search offerings through browsing, and if workers find a preferred offering through this channel, they may subsequently favorite the underlying template from which that offering was generated.}

Figure~\ref{fig:flow_from_recommendation_to_match} summarizes how template recommendations translate into realized matches through the favorite list. The process has two stages. First, workers build their favorite lists by opening the template recommendation tab, viewing personalized template recommendations, and adding templates of interest. Second, the accumulated favorite list serves as a channel through which workers learn about new offerings posted from those templates and apply to those offerings when suitable opportunities are available. Thus, recommendations affect matching not only through immediate user responses, but also by gradually shaping the stock of templates from which workers receive future opportunities. This cumulative feature is central to our analysis of favorite-list management.

\subsection{User Interface}\label{sec:user_interface}

\begin{figure}[!tp]
    \begin{center}
        \begin{subfigure}{0.45\textwidth}
            \centering
            \includegraphics[width=\textwidth]{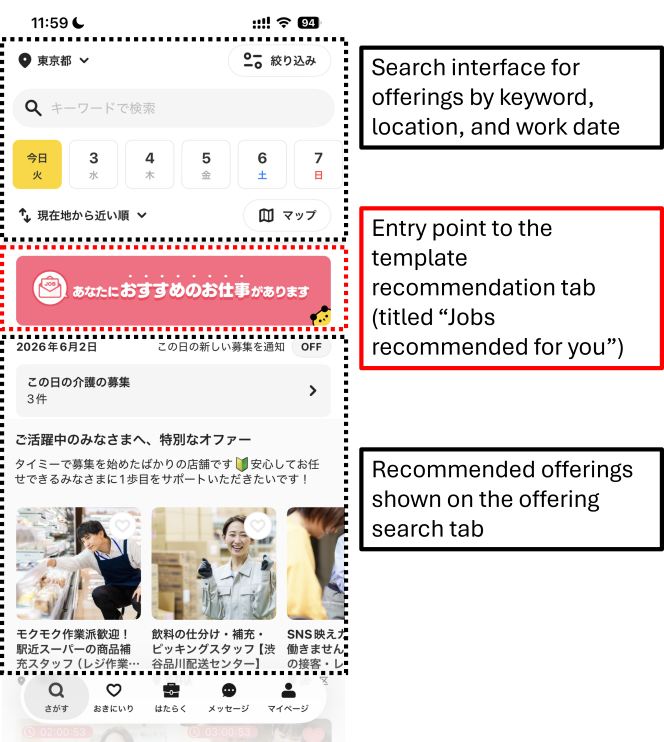}
            \caption{Offering Search Tab}
            \label{fig:offering_search_tab}
        \end{subfigure}
        \hfill
        \begin{subfigure}{0.45\textwidth}
            \centering
            \includegraphics[width=\textwidth]{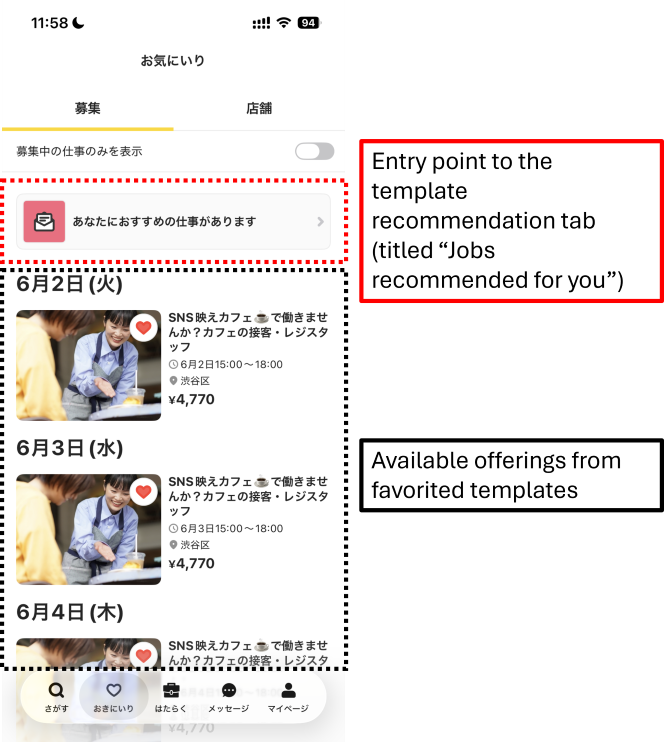}
            \caption{Favorite List Tab}
            \label{fig:favorite_list_tab}
        \end{subfigure}\\
        \medskip
        \begin{subfigure}{0.45\textwidth}
            \centering
            \includegraphics[width=\textwidth]{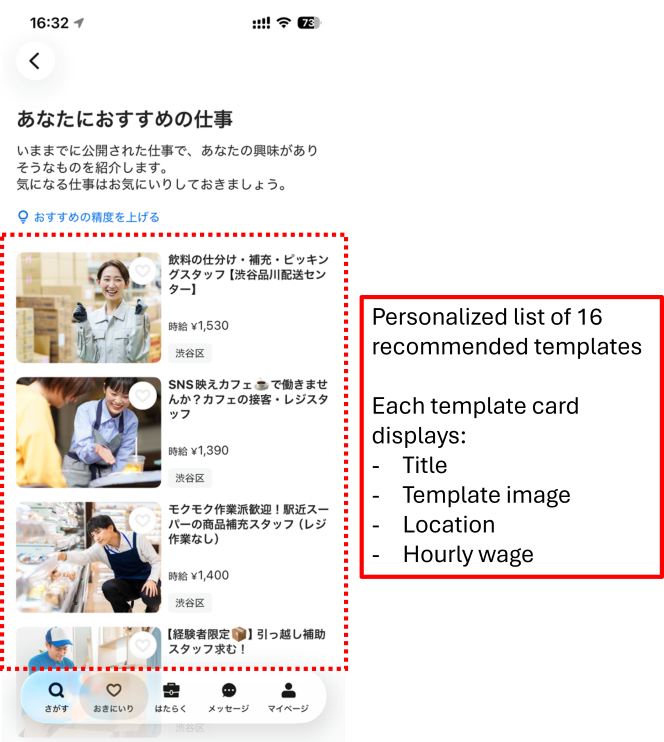}
            \caption{Template Recommendation Tab}
            \label{fig:template_recommendation}
        \end{subfigure}
        \hfill
        \begin{subfigure}{0.45\textwidth}
            \centering
            \includegraphics[width=\textwidth]{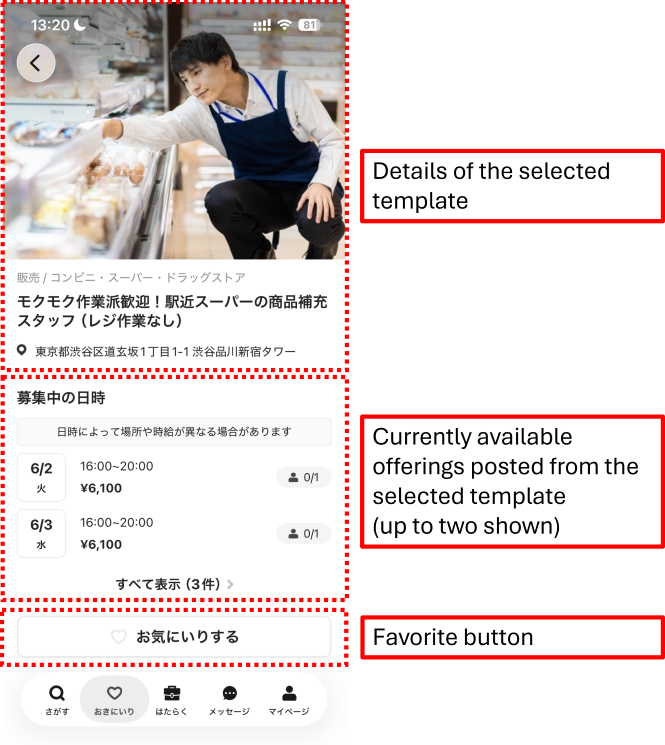}
            \caption{Template Detail Tab}
            \label{fig:favorite_list_post_click_tab}
        \end{subfigure}
        \caption{Template Recommendation Flow in the Timee App}
        \label{fig:app_screenshot}
    \end{center}
    \footnotesize
    Note: Tapping the pink-highlighted tab with the mailbox icon in panels~(a) or~(b) navigates to the template recommendation tab~(c). Panel~(c) displays a ranked list of recommended templates. Clicking a recommendation card opens the template detail tab~(d), where the worker decides whether to add the template to their favorite list. If the template has unfilled offerings, up to two currently open postings are shown at the bottom of~(d), providing the worker with information on the template's unfilled capacity. The figures are mockups based on the user interface of the Timee app rather than actual screenshots. All templates and offerings shown in the figures, including card images, addresses, wages, and other information, are fictional and do not correspond to any actual template or offering on the platform. All of the card images are licensed from Adobe Stock.
\end{figure}

Figure~\ref{fig:app_screenshot} illustrates the recommendation flow through the app's user interface. Workers can reach the template recommendation tab (Figure~\ref{fig:template_recommendation}) from either the offering search tab (Figure~\ref{fig:offering_search_tab}) or the favorite list tab (Figure~\ref{fig:favorite_list_tab}) by clicking the highlighted banner. The recommendation tab displays 16 templates, refreshed every hour. For each template, the interface shows a profile image selected by the firm, the template title (often accompanied by a brief promotional description), the hourly wage, and the approximate location. Clicking a recommended template opens the template detail tab (Figure~\ref{fig:favorite_list_post_click_tab}), where the worker views detailed information, including job tasks, required skills, and necessary items. If the template has unfilled offerings, up to two currently open postings are displayed at the bottom, providing the worker with direct information on the template's unmet labor demand. Workers then decide whether to add the template to their favorites. Once a template is added to a worker's favorite list, the worker receives push notifications when new offerings are posted from that template, and those offerings also appear in the worker's favorite list tab (Figure~\ref{fig:favorite_list_tab}).

\subsection{Current Recommender}\label{subsec:points_for_improvement}

Timee currently employs a recommender that selects and displays the top 16 templates with the highest predicted probability of being favorited for each worker. We refer to this approach as the \emph{greedy recommender} (\Greedy) in this paper. These predictions are generated using machine-learning technology; however, the specific prediction methodology is not the focus of this paper and is therefore not described in detail. The underlying design principle of the current system---choosing recommendations to maximize the probability of favoriting---is a standard and widely adopted approach for favoriting-oriented recommender features.

Nevertheless, \Greedy is inefficient in at least three important respects when the objective is to improve realized matching outcomes. First, \Greedy concentrates exposure on a small subset of templates. Predicted favoriting probabilities are often positively correlated across workers, so popular templates are repeatedly selected by \Greedy. For example, in the baseline simulation calibrated to 2024 Hokkaido data, the top 10\% of templates by recommendation count account for 36.7\% of all recommendations. This concentration can be problematic in spot-work markets, where insufficient exposure directly increases the risk that offerings associated with less popular templates remain unfilled. From the worker's perspective, concentrated recommendations also reduce the value of the favorite list: even when offerings are posted by favorited templates, they may be taken by other workers before the worker can apply, limiting access to viable opportunities.

Second, \Greedy does not take into account how many subscribers a template has already accumulated. Favoriting a template leads to subscribers receiving notifications when new offerings are posted, from which some workers may apply, ultimately contributing to the fulfillment of job postings. As a result, even if two templates have identical favoriting probabilities, allocating more exposure to the template with fewer existing subscribers (for instance, because it is newly created) can be more effective in increasing the likelihood that future offerings are filled. By ignoring the current size of the subscriber base, the existing recommender system fails to allocate exposure in a way that reflects diminishing returns to additional subscribers.

Third, \Greedy does not account for heterogeneity in labor demand, that is, differences in how frequently templates generate offerings and how many workers they hire per offering. In practice, labor demand varies substantially across firms. Some templates generate dozens of offerings every day, while others may not post a single offering for weeks. While Timee excludes completely inactive templates that have not posted any offerings recently, there remains significant variation in labor demand across ``active'' templates. Naturally, templates with greater labor demand (i.e., those that frequently generate a large volume of offerings) require more exposure; however, these templates do not necessarily have high favoriting probabilities.

These three features of \Greedy give rise to what we refer to as misdirected concentration of exposure, which adversely affects the welfare of both workers and firms. Even if a worker favorites a template that is popular but has limited labor demand, they are unlikely to obtain meaningful employment opportunities from it: the number of offerings is small, and those offerings are quickly accepted by the large pool of existing subscribers. An analogy from e-commerce would be a recommender system that promotes attractive items that tend to be out of stock. Users may like recommended items, but the recommendation creates limited purchasing opportunities.

From the firms' perspective, misdirected concentration is also inefficient. Templates that receive little exposure face a higher risk of unfilled offerings, even when they represent substantial labor demand. At the same time, additional exposure to already highly subscribed templates generates no incremental benefit once offerings are almost surely filled. Even if many workers apply to an offering, the firm cannot choose among them, and the offering is effectively filled on an FCFS basis. Consequently, excessive concentration of subscribers on certain templates provides no additional benefit to firms, while depriving other templates of the exposure needed to fill their offerings.

To mitigate misdirected concentration, a recommender system should place upper bounds on the exposure of templates that are favored by many workers, while allocating more exposure to templates with fewer subscribers or larger labor demand. However, determining how to weight these factors and how to optimally set exposure levels is a challenging problem. The optimal policy depends sensitively on a range of parameters that characterize spot-work markets, including the distribution of worker preferences, the heterogeneity of labor demand, and the dynamics of offerings and acceptance. Moreover, the feedback effects induced by changes in recommendation policies---through their impact on the distribution of subscribers and, in turn, on application behavior---are highly complex and not amenable to tractable analytical characterization. For these reasons, rather than deriving an optimal recommender system from a simple and abstract analytical model, we evaluate recommender performance using simulations based on a rich model that closely mirrors the matching process on Timee, and we complement these analyses with field experiments to demonstrate the effectiveness of our proposed approach in practice.

\section{Recommender Design}\label{sec:recommender_design}

In this section, we describe the current recommender (\Greedy) and the three alternatives, static quota (\SQ), adaptive quota (\AQ), and thresholded eligibility control (\TEC). 

\subsection{Greedy}
The \emph{greedy recommender} (\Greedy) is designed to mimic Timee's current recommender. For each worker, the system considers templates that the worker has not yet added to their favorite list and greedily recommends the top $c$ templates with the highest predicted favoriting probabilities, where $c = 16$ in the current implementation.

As discussed in Section~\ref{subsec:points_for_improvement}, \Greedy tends to overly concentrate exposure on popular templates, while underexposing templates with fewer subscribers or larger labor demand. This behavior leads to misdirected concentration. The three recommenders introduced below are designed to address these shortcomings.

\subsection{Static Quota (SQ)}
The \emph{static quota recommender} (\SQ) prevents concentration by imposing a uniform upper bound on the number of workers to whom each template can be recommended. Specifically, each template is assigned a fixed quota $q$, which is treated as an external parameter. Under this constraint, the recommendation problem can be viewed as a many-to-many matching problem between workers, each with recommendation capacity $c = 16$, and templates, each with capacity $q$. To generate fair and efficient recommendations satisfying these capacity constraints, we employ a round-robin variant of \emph{random serial dictatorship} (RSD). That is, workers are randomly ordered, and each worker sequentially selects their most preferred template, defined as the template with the highest favoriting probability for that worker, from the set of templates with remaining quota. After all workers have selected one template, the process repeats in the same order until each worker's recommendation list reaches capacity $c$. Note that \Greedy can be interpreted as \SQ with $q = +\infty$.

While \SQ\ equalizes exposure across templates, it induces exposure misallocation because it does not account for heterogeneity in either the current number of subscribers or labor demand. A more efficient quota policy would adjust template-level quotas according to each template's current demand for exposure, assigning higher limits to templates with relatively few subscribers or greater expected labor demand.

However, implementing such differentiated quotas raises both fairness and practical concerns. Assigning template-specific quotas requires the platform to make discretionary judgments across templates, which may be perceived as arbitrary or inequitable. Moreover, even if one attempts to base quotas on observable categories such as region or job type, defining appropriate boundaries for such categories is inherently difficult and may introduce further distortions. As a result, tuning these quotas ex ante is difficult in practice.

\subsection{Adaptive Quota (AQ)}
The \emph{adaptive quota recommender} (\AQ) addresses \SQ's shortcomings by automatically adjusting individual quotas based on each template's recent market activity and performance. Specifically, \AQ updates individual quotas using information on (i) how actively a template has posted offerings in the recent past and (ii) whether those offerings were successfully filled.

Three values parameterize \AQ: (i) the average quota $\bar{q}$, (ii) a score weight $w^0$ assigned per unit of posted offering capacity, and (iii) a score weight $w^1$ assigned per unit of offering capacity that remains unfilled. At the beginning of each round, \AQ resets all template scores to zero. If template $k$ posts $x_k^0$ units of offering capacity and $x_k^1$ units of that capacity remain unfilled in the previous round, its score for that round is given by $p_k = w^0 x_k^0 + w^1 x_k^1$. \AQ maintains the specified average quota $\bar{q}$ while allocating individual quotas across templates in proportion to their scores. That is,
\begin{equation}
    q_k = \bar{q} |K| \frac{p_k}{\sum_{k' \in K} p_{k'}}.\label{eq:adaptive_quota}
\end{equation}
Finally, in an environment where each worker has a uniform capacity $c$ and each template $k$ has capacity $q_k$ specified by \eqref{eq:adaptive_quota}, recommendations are generated using a round-robin type of RSD, as in \SQ.

In \AQ, templates that are actively posting offerings and exhibiting large unfilled labor demand (captured by higher values of $x_k^0$ and $x_k^1$) receive larger quotas and are therefore prioritized in recommendations. At the same time, this prioritization is determined entirely by observable market activity and performance, rather than by discretionary, template-specific decisions made by the platform. This feature mitigates fairness concerns associated with manual differentiation across templates.

Although \AQ addresses the key inefficiencies of \SQ, it is too slow to deploy on a large platform such as Timee because it generates recommendations using a round-robin variant of RSD. In RSD, workers select templates one by one, and the set of templates available to a worker depends on the selections made by earlier workers. Therefore, the platform must continuously update remaining template quotas while constructing recommendation lists. This sequential dependence prevents the computation from being parallelized across workers. It also requires the platform to maintain a large global state of remaining quotas and worker-template favoriting probabilities during the recommendation process, creating substantial memory pressure.

\subsection{Thresholded Eligibility Control (TEC)}
\label{subsec:tec}

\subsubsection{Core Idea}

\TEC is designed to approximate the exposure-allocation logic of \AQ while avoiding a sequential implementation. The key idea of \TEC is to replace explicit quotas with \emph{eligibility thresholds}. In \AQ, a template with a larger quota remains available for more selections in the round-robin RSD procedure. In \TEC, the same idea is implemented by assigning each template an eligibility threshold. A template with a larger threshold remains eligible at later selection timings.

At the beginning of each round, \TEC assigns each template a \emph{score}. Parallel to \AQ, the score represents the template's current demand for exposure. Templates with more recent posting activity or more unfilled capacity accumulate larger scores through the update rule. These scores are then directly converted into eligibility thresholds $\tau_{k}$. Each worker $i \in I$ and each selection step $s = 1, \dots, c$ in the recommendation list is assigned a selection timing $z_{is}$. A template $k \in K$ is eligible for that selection if its threshold $\tau_k$ is larger than the selection timing $z_{is}$. In this manner, \TEC specifies the ``choice set'' (the set of templates eligible for recommendation) without running round-robin RSD.

This construction mimics the logic of round-robin RSD. Early selections face a larger choice set, while later selections face a smaller choice set that contains only templates with larger exposure demand. The important difference is that eligibility depends only on precomputed eligibility thresholds and not on recommendations made to other workers. Therefore, once thresholds are computed, each worker's recommendation list can be constructed independently and in parallel.

The construction has two additional components. First, before computing thresholds, \TEC caps scores so that no single template receives more than one slot's worth of eligibility mass. This prevents the threshold rule from creating empty choice sets for purely mechanical reasons. Second, if the choice set is still empty for a worker and a selection position, \TEC completes the recommendation list using the same rule as \Greedy.

\subsubsection{Scores and Eligibility Thresholds}

At the beginning of each round, \TEC takes as input a nonnegative score $p_k$ for each template $k \in K$. This score is the state variable used to determine the template's exposure priority in the current round.\footnote{The carried-over score of a template may become negative after the score update. When computing eligibility thresholds, we truncate such scores at zero and use $\max\{p_k,0\}$. This truncation affects only the threshold calculation.} A higher score means that the template should remain eligible for recommendation for a larger part of the round. Let $P = \sum_{k \in K} p_k$ denote the total score in the current round. As described below, these scores are updated in each round according to a score update rule, while their total is maintained at a constant level, $P$.

The eligibility thresholds are constructed from these scores. More precisely, before the scores are used as inputs to the algorithm, we apply a procedure called score capping, detailed in Appendix~\ref{subsec:score_capping}, which limits extremely large scores. We first order templates by their current scores: $p_{\pi(1)} \ge p_{\pi(2)} \ge \cdots \ge p_{\pi(|K|)}$. Ties are broken by an arbitrary fixed rule. For each $j = 1, \dots, |K|$, we define the eligibility threshold of template $\pi(j)$ by
\begin{equation}
    \tau_{\pi(j)}
    =
    \frac{1}{P}\sum_{h=j}^{|K|} p_{\pi(h)},
    \text{ for }j=1,\ldots,|K|.
\end{equation}
For notational convenience, we define $\tau_{\pi(|K|+1)} = 0$. 




\subsubsection{Recommendation and Completion}

At the beginning of each round, the platform assigns each worker a unique integer $i \in \{1,\ldots,|I|\}$ uniformly at random. We refer to the worker assigned integer $i$ as worker $i$. For worker $i$, let $\mathcal K_i \subseteq K$ denote the set of templates considered as recommendation candidates. Specifically, $\mathcal{K}_i$ consists of templates that satisfy certain eligibility requirements, such as being sufficiently close to worker $i$'s location, and that worker $i$ has not yet added to their favorite list. Let $R_{i,s-1}$ be the set of templates already added to worker $i$'s recommendation list before selection step $s$ in the current round.

For worker $i$ and selection step $s=1,\ldots,c$, define the selection timing $z_{is}$ as follows:
\begin{equation}
    z_{is}
    =
    \frac{1}{c}
    \left(
        s-1+\frac{i-1}{|I|}
    \right).
\end{equation}
The term $s-1$ captures how many recommendations have already been selected for the worker. The term $(i-1)/|I|$ captures the worker's position in the random ordering. As in round-robin RSD, the selection step $s$ is the primary determinant of timing, while the random worker order refines the timing within each step.

The choice set is $\mathcal{C}_{is} = \{k \in \mathcal K_i \setminus R_{i,s-1}:\tau_k \ge z_{is}\}$.
If $\mathcal{C}_{is}$ is nonempty, \TEC recommends the template with the highest predicted favoriting probability among templates in $\mathcal{C}_{is}$.

The choice set $\mathcal{C}_{is}$ may be empty. In particular, in the final selection step ($s = c$), many workers' choice sets consist of only a small number of templates. If a worker has already selected all of these templates in earlier selection steps, then $\mathcal{C}_{is}$ becomes empty. When $\mathcal{C}_{is}$ is empty, \TEC uses a completion rule. The completion rule fills the remaining position in the same way as \Greedy: it ignores eligibility thresholds and chooses the best remaining feasible template for the worker. Specifically, \TEC recommends the template with the highest predicted favoriting probability among templates in $\mathcal{K}_i \setminus R_{i,s-1}$. This completion rule is used only as a fallback. The main role of \TEC is to control exposure through eligibility thresholds. The completion rule ensures that the recommendation list is still filled whenever feasible templates remain, even when the thresholded rule cannot identify a feasible candidate.

\subsubsection{Score Update}

The crucial difference between \TEC and round-robin RSD concerns how template eligibility evolves when a template is not actually recommended. In round-robin RSD, a template's quota is consumed only when the template is recommended. In contrast, \TEC determines eligibility from the selection timing $z_{is}$ and the precomputed threshold $\tau_k$. Therefore, a template may lose eligibility as the timing advances even if it has received no recommendation.

This issue is important for templates with low predicted favoriting probabilities. Such templates may have positive labor demand and therefore positive scores, but they may still fail to be recommended if other eligible templates have higher predicted favoriting probabilities. If scores were reset every round as in \AQ, these templates could continue to receive little exposure even though their labor demand is high and they keep earning high scores. 

To avoid this problem, \TEC carries scores across rounds and subtracts scores only according to realized exposure. At the end of each round, \TEC first computes a positive score increment $p_k^+$ for each template:
\begin{equation}
    p_k^+
    =
    w^0 x_k^0 + w^1 x_k^1,
\end{equation}
where $x_k^0$ is the posted offering capacity of template $k$ in the round, $x_k^1$ is the unfilled capacity of template $k$ in the round, and $w^0$ and $w^1$ are nonnegative score weights.

Next, \TEC computes a score reduction based on realized recommendation exposure. Let $r_k$ denote the number of recommendations received by template $k$ in the round. Define
\begin{equation}
    p_k^-
    =
    \frac{r_k}{\sum_{l \in K} r_l}
    \sum_{l \in K} p_l^+.
\end{equation}
The score carried into the next round is
\begin{equation}
    p'_k
    =
    p_k + p_k^+ - p_k^-.
\end{equation}
Note that this update rule preserves the total score: $\sum_{k \in K}p_k' = \sum_{k \in K}p_k = P$.

The interpretation is straightforward. If a template posts many offerings or has large unfilled capacity, it receives a large positive increment $p_k^+$. If it also receives many recommendations, part of this increment is offset by $p_k^-$. If it receives little exposure relative to its market need, its score increases and it remains eligible for a larger part of future rounds. Conversely, if it receives substantial exposure, its score decreases. In this way, \TEC preserves the quota-consumption intuition of \AQ: exposure demand is reduced only when exposure is actually delivered. At the same time, because recommendation lists are generated from precomputed thresholds, the algorithm remains fully parallelizable.

\section{Simulation Study}\label{sec:simulation}
\subsection{Model}\label{subsec:model}

This section describes an economic model used in our simulation study. To enable market size to vary in the simulations, we define worker and template types based on the real data. We then specify the primitives of the simulation as functions of these types and their combinations, including the probability that a worker favorites a template when it is recommended and the labor demand generated by each template. After initialization, the simulation proceeds through repeated rounds, each consisting of a sequence of four distinct phases: replacement, recommendation, favoriting, and matching.

\paragraph{Initialization}
We introduce one template of each type $\theta(k)$ from the set $\Theta_K$ into the market. Additionally, we introduce $I$ workers into the market. Each worker starts with an empty favorite list and is assigned a type $\theta(i) \in \Theta_I$ drawn uniformly at random. A worker's type $\theta(i)$ determines the probability that they favorite each template.

\paragraph{Replacement Phase}
A fraction $\delta \in (0,1)$ of workers exit the market, and the same number of new workers enter the market. Each new worker starts with an empty favorite list and a type $\theta(i)$ drawn uniformly at random from the set $\Theta_I$.

\paragraph{Recommendation Phase}
Templates are recommended to each worker according to a recommender (\Greedy, \SQ, \AQ, or \TEC). For each worker, the recommender displays up to $c$ templates. This setting corresponds to Timee's current policy of displaying 16 templates to each worker at a time, and we do not allow alternative recommenders to increase the number of displayed templates.

\paragraph{Favoriting Phase}
Each worker views their recommendation list with probability $\gamma \in (0, 1)$. If a worker $i$ views the list, they favorite each template $k$ from the list independently, with probability $\alpha_{\theta(i) \theta(k)} \in [0, 1]$.

\paragraph{Matching Phase}
Each template draws the number of offerings for the round from a distribution $F_{\theta(k)}$, i.i.d. Each worker searches for offerings with probability $\lambda \in (0, 1)$. Workers who actively search are randomly ordered, and each worker applies to the ``most preferred'' template among those that (i) have been added to their favorite list and (ii) have an open offering. An open offering is defined as an offering posted in the current round whose capacity has not yet been fully filled by workers who searched earlier (in terms of the random order drawn in the current matching phase). In the simulation, each worker's ``preference'' over templates is proxied by their favoriting probabilities when recommended. That is, a larger value of $\alpha_{\theta(i) \theta(k)}$ indicates that worker $i$ prefers template $k$ more. If none of the templates in the worker's favorite list have any open offerings, the worker does not apply and becomes unemployed in that round.

\medskip

The model described in this section, together with the specification of the recommender system in Section~\ref{sec:recommender_design}, induces a Markov process on the state space of workers' favorite list profiles and persistent types, as well as recommender-specific state variables such as scores.

To evaluate recommender performance with a realistic simulation setting, we select parameters to closely replicate the characteristics of Timee's spot work market in Hokkaido during 2024, using real-world data. Hokkaido is well-suited for this simulation exercise because it is geographically separated from the rest of Japan, allowing us to abstract from interactions with spot-work markets in other regions. Specifically, we sample workers and templates that were active in Hokkaido during the targeted months and predict the favoriting probabilities $\alpha_{\theta(i) \theta(k)}$ using Timee's latest prediction model. We further set parameters $F_{\theta(k)}$, $\gamma$, and $\lambda$ based on the empirical frequencies with which templates posted offerings and workers used Timee's recommendation and matching features. We sampled 838 templates active in Hokkaido during the targeted period. We fix the total number of workers at 1{,}000 to ensure that the average number of offerings posted by these templates roughly matches the number of active workers in each round.

\subsection{Setup}

We analyze the long-run market configuration induced by each recommender once the favorite lists have adjusted and the economy has approximately reached a steady state. These long-run outcomes reveal how the recommender changes exposure allocation, subscriber stocks, and employment when operated for a sufficiently long period. The recommender's performance in the steady state is therefore the primary criterion for policy evaluation.

We compare economies in which a given recommender (\Greedy, \SQ, \AQ, or \TEC) is used from the beginning of the simulation. As in the baseline environment, all workers start with empty favorite lists, and thus the stock of favorites accumulates endogenously over time. We regard the economy as having reached an approximate steady state when the moving average of the total number of favorites over the most recent 15 rounds no longer exceeds that over the preceding 15 rounds. After that point, we record outcomes for the subsequent 50 rounds. To account for simulation variability, we repeat this process for 100 independent sample paths. The results reported in the subsequent sections represent the averages across these 100 paths. This exercise is designed to answer an eventual-effect question: if the platform were to operate each recommender for a sufficiently long period, what market-level allocation of exposure and what level of employment would ultimately emerge?


For \SQ, we set the quota parameter $q = 25$. For \AQ and \TEC, we adopt the average quota $\bar{q} = 40$, the score weight for posting $w^0 = 50$, and the score weight for unfilled capacity $w^1 = 125$. These parameters were the best for the baseline simulation setting among those we tested. Further tuning could enhance performance even more. Nevertheless, the robustness analysis in Section~\ref{subsec:simulation_robustness} shows that the main qualitative comparison across recommenders is not driven by this particular baseline calibration.

\subsection{Results}

\subsubsection{Job-Finding Rate and Fill Rate}

\begin{table}[!t]
  \begin{center}
     \caption{Per-Round Job-Finding Rate and Fill Rate}
     \begin{tabular}[t]{lcc}
\toprule
  & Per-Round Job-Finding Rate & Per-Round Fill Rate \\
\midrule
Greedy & 0.5761 (0.0067) & 0.6742 (0.0081) \\
SQ & 0.6420 (0.0054) & 0.7525 (0.0087) \\
AQ & 0.6953 (0.0064) & 0.8144 (0.0068) \\
TEC & 0.7003 (0.0061) & 0.8217 (0.0088) \\
\bottomrule
\end{tabular}
    \label{tab:job_finding_fill_rate_table}  
  \end{center}
    \footnotesize
  \textit{Note}: The reported values represent the average of each metric over 50 rounds in each simulation, averaged across 100 independent simulation paths. The numbers in parentheses indicate the standard deviations across the 100 paths.
\end{table}

First, we examine the per-round job-finding rate, defined as the fraction of workers actively searching for offerings in the matching phase of that round who successfully apply to an offering. 

Table~\ref{tab:job_finding_fill_rate_table} summarizes (i) the mean per-round job-finding rate, the share of workers actively searching offerings in a round who successfully match with an offering, and (ii) the mean per-round fill rate, the share of offerings posted in a round that successfully match with a worker. These two ratios have the same numerator: the number of matches formed between workers and offerings in that round. Their denominators are, respectively, the number of active workers and the number of posted offerings in that round, both of which are determined exogenously with respect to the recommender design. Therefore, these two measures are approximately aligned.

We find that the per-round job-finding rate reached 57.6\% under \Greedy, improved to 64.2\% with \SQ, and further increased to 69.5\% with \AQ and 70.0\% with \TEC. The per-round fill rate exhibits the same ordering, rising from 67.4\% under \Greedy to 75.3\% under \SQ, 81.4\% under \AQ, and 82.2\% under \TEC. These results indicate that exposure control improves matching outcomes on both sides of the market: active workers are more likely to find offerings, and posted offerings are more likely to be filled.

Note that, in our simulation, the population ratio between workers and templates was chosen arbitrarily; therefore, the per-round job-finding rate differs from the actual rate realized on Timee. In practice, workers also favorite templates or accept offerings through other channels, such as search, and thus precisely measuring the contribution of recommendation alone to filling offerings is challenging. The important point is that our proposed methods (\AQ, \TEC) clearly outperform the existing recommender (\Greedy), as well as a simple quota policy (\SQ).

\begin{figure}[!tp]
  \centering
  \begin{subfigure}{0.8\textwidth}
    \centering
    \includegraphics[width=\textwidth]{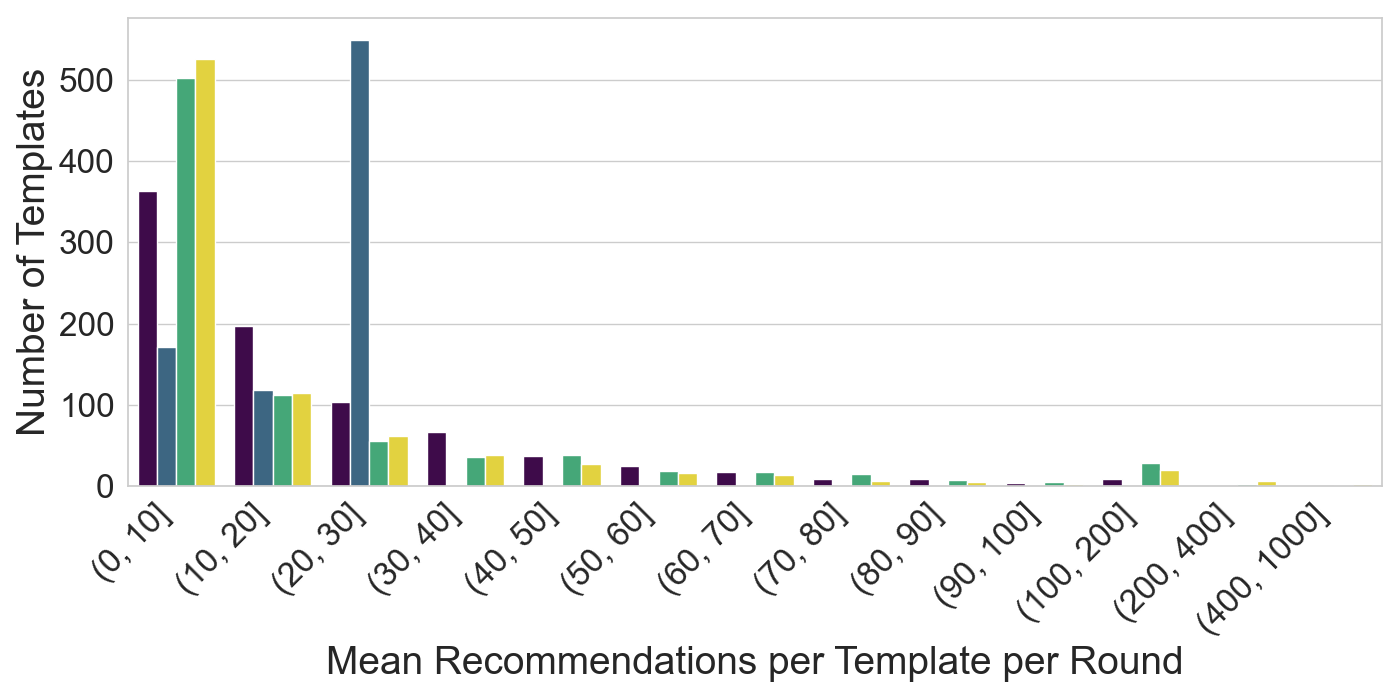}
    \caption{Distribution of Recommendation Counts per Round}
    \label{fig:distribution_of_recommendations}
  \end{subfigure}\\
  \begin{subfigure}{0.8\textwidth}
    \centering
    \includegraphics[width=\textwidth]{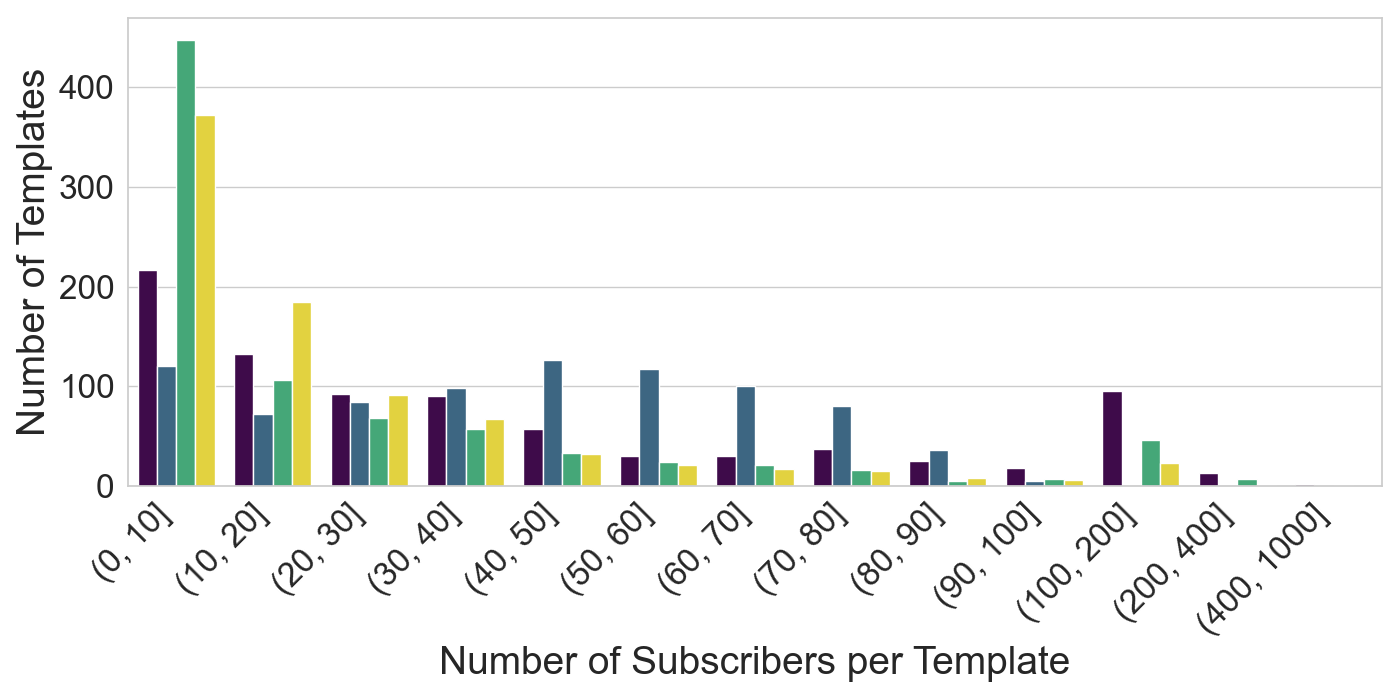}
    \caption{Distribution of the Number of Subscribers}
    \label{fig:distribution_of_subscribers}
  \end{subfigure}\\
  \includegraphics[width=0.45\textwidth]{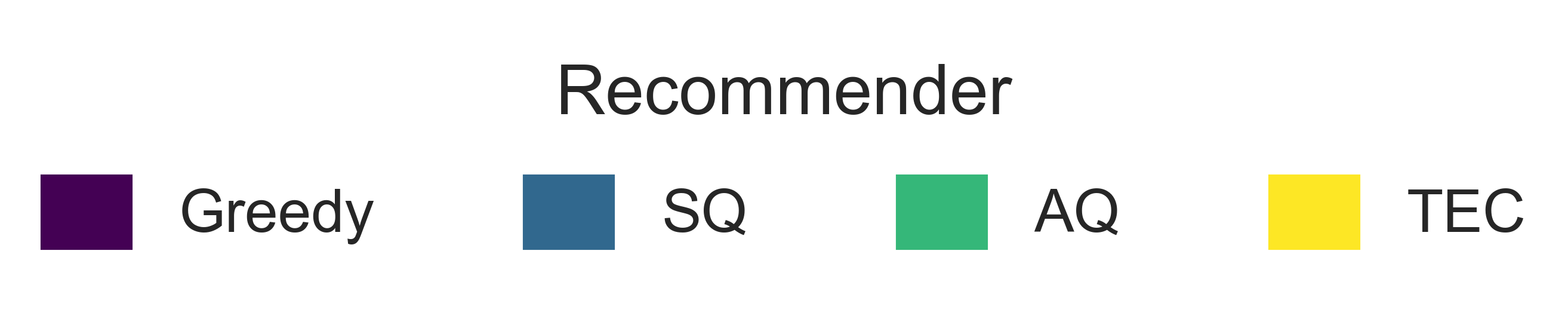}
  \caption{Template-Level Exposure and Subscriber Distributions in the Baseline Simulation}
  \label{fig:simulation_template_outcomes}
\end{figure}

\subsubsection{Template-Level Outcome Distributions} Figure~\ref{fig:simulation_template_outcomes} illustrates the distribution of times each template is recommended (Figure~\ref{fig:distribution_of_recommendations}) and the distribution of the number of subscribers, i.e., workers who favorited that template (Figure~\ref{fig:distribution_of_subscribers}). Under \Greedy, both distributions exhibit heavy tails: some templates are recommended to a large number of workers, while many others receive little to no exposure. As a consequence, there is substantial dispersion in the number of subscribers across templates. This pattern reflects the fact that templates that are highly likely to be favorited by many workers are repeatedly recommended under \Greedy. In contrast, under \SQ, the number of recommendations each template can receive in a round is capped at 25, which leads many templates to receive a similar number of recommendations. As a result, the distribution of subscribers across templates is considerably more balanced.

In contrast, \AQ and \TEC induce even more pronounced concentration in recommendations and subscribers than \Greedy. This pattern arises because both \AQ and \TEC are designed to prioritize templates that post a large volume of offerings and experience unfilled capacity, thereby allocating more recommendations to such templates. Importantly, while \Greedy, \AQ, and \TEC all generate concentration in exposure, the nature of this concentration differs.

\begin{figure}[!tp]
    \begin{center}
    \begin{subfigure}{\textwidth}
    \centering
    \includegraphics[width=0.65\textwidth]{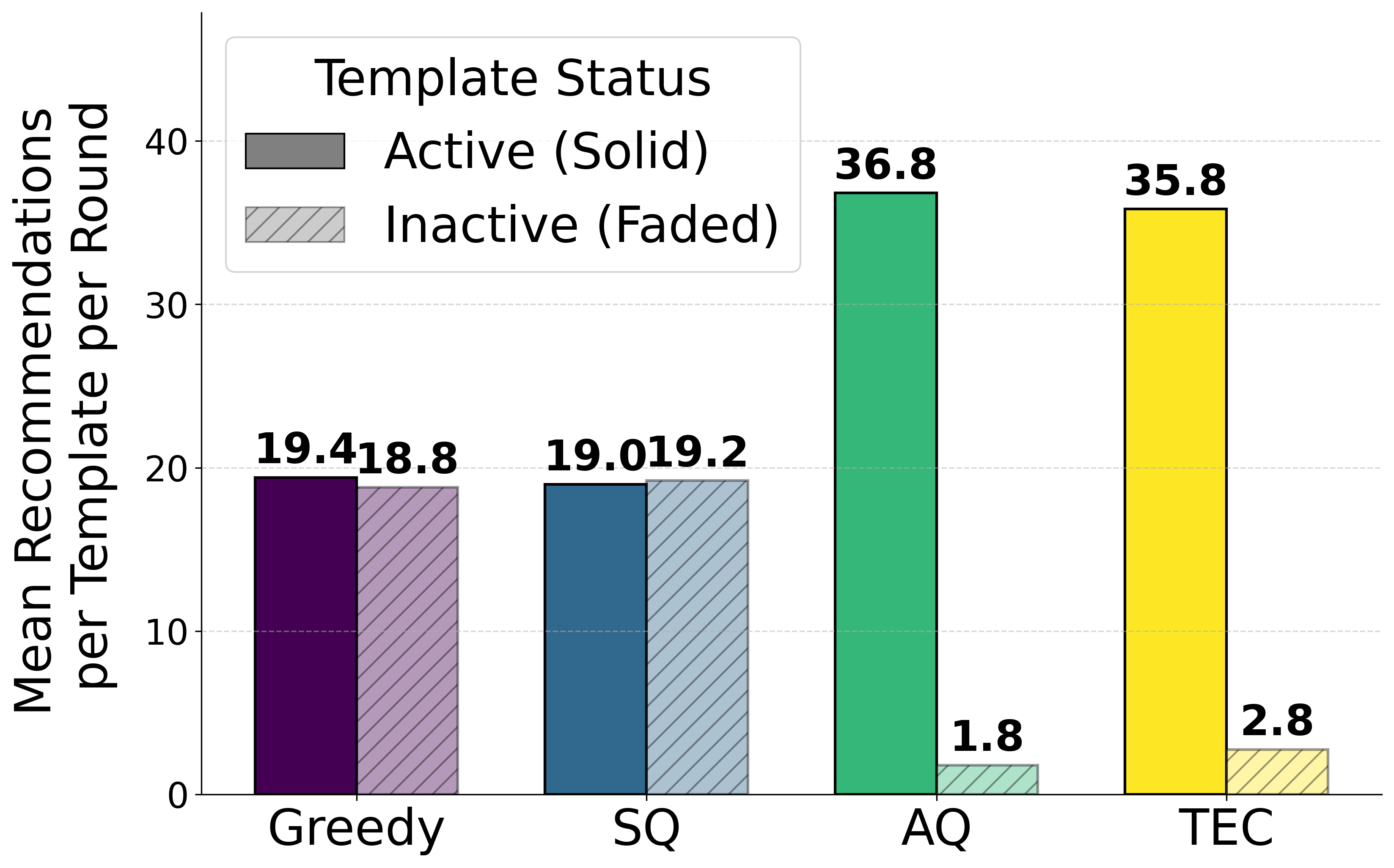}
    \caption{Exposure by Template Activity Status}
    \label{fig:simulation_capacity_bin_active_vs_inactive}
    \end{subfigure}
    \\
    \begin{subfigure}{\textwidth}
    \centering
    \includegraphics[width=0.65\textwidth]{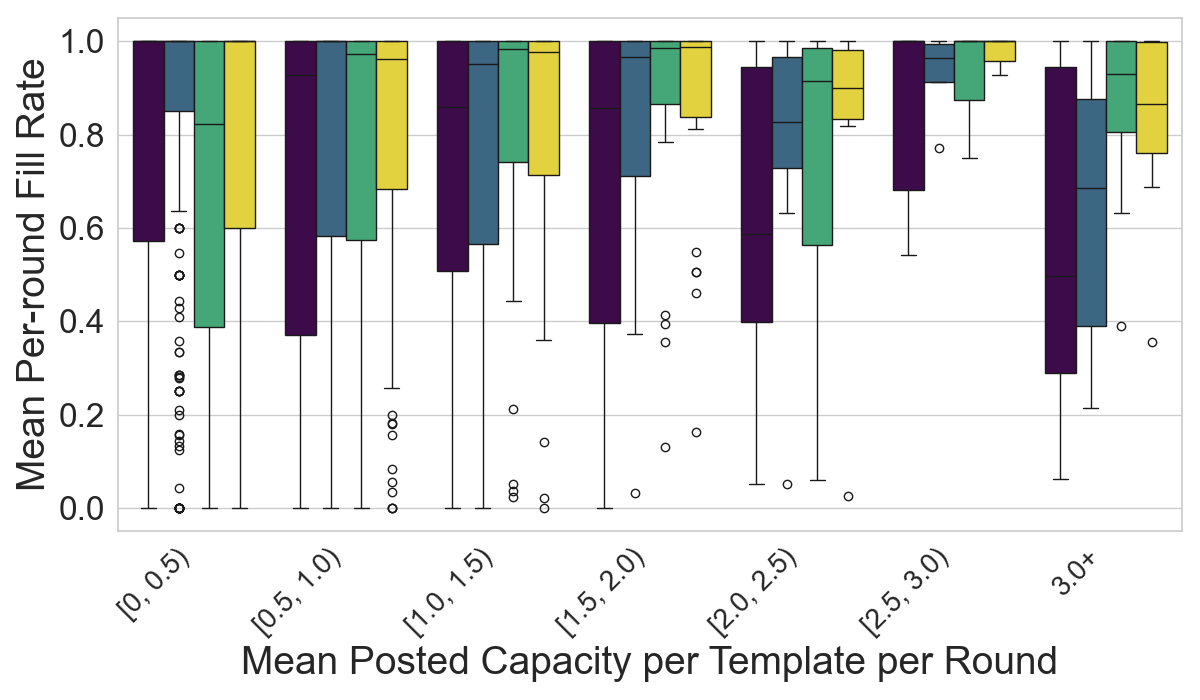}
    \caption{Per-Round Fill Rate by Posted Capacity}
    \label{fig:simulation_capacity_bin_across}
    \end{subfigure}
    \includegraphics[width=0.45\textwidth]{figuretable/recommendation_cap_adjustment_project/ec_graph/legend_only.png}
    \caption{Exposure Allocation and Per-Round Fill Rates by Template Activity}
    \end{center}
    \footnotesize
    Note: ``Active templates'' in Figure~\ref{fig:simulation_capacity_bin_active_vs_inactive} are those with mean posted capacity of at least 0.1 per round; inactive templates are all others.
  \label{fig:simulation_capacity_bin}
\end{figure}

\subsubsection{Concentration on Active Templates} Figure~\ref{fig:simulation_capacity_bin_active_vs_inactive} visualizes how \AQ and \TEC allocate exposure toward templates with greater labor demand. For exposition, we classify templates that post fewer than 0.1 offerings per round on average as inactive templates, and refer to all remaining templates as active templates. Under \Greedy, exposure is concentrated on templates with high predicted favoriting probabilities; however, because template popularity is not necessarily correlated with labor demand, \Greedy allocates a similar number of recommendations to active and inactive templates. Likewise, \SQ is explicitly designed to prevent concentration and therefore recommends active and inactive templates at comparable rates. In both cases, \Greedy and \SQ recommend active and inactive templates approximately 19 times per round, on average. In contrast, \AQ and \TEC generate substantial differences in exposure across template types. Under \AQ, active templates are recommended an average of 36.8 times per round, while inactive templates are recommended only 1.8 times; under \TEC, the corresponding figures are 35.8 and 2.8. These results illustrate that \AQ and \TEC deliberately direct exposure toward templates with higher labor demand.

Figure~\ref{fig:simulation_capacity_bin_across} shows templates' per-round fill rates, defined as the ratio of accepted offering capacity to total offered capacity in each round. To highlight variations in activity levels across templates, templates are grouped into bins according to their per-round average offering capacity. \SQ consistently outperforms \Greedy across nearly all capacities. In contrast, since \AQ and \TEC prioritize templates with greater labor demand, their per-round fill rate is comparatively lower for smaller-capacity templates but significantly higher for larger-capacity templates. Notably, among the four recommenders, only \AQ and \TEC effectively meet the labor demands of large-capacity templates, which explains why \AQ and \TEC achieve the highest per-round job-finding rate and fill rate at the aggregate level.

\subsubsection{Robustness}\label{subsec:simulation_robustness}
\begin{figure}[!tp]
    \centering
    \includegraphics[width=0.9\textwidth]{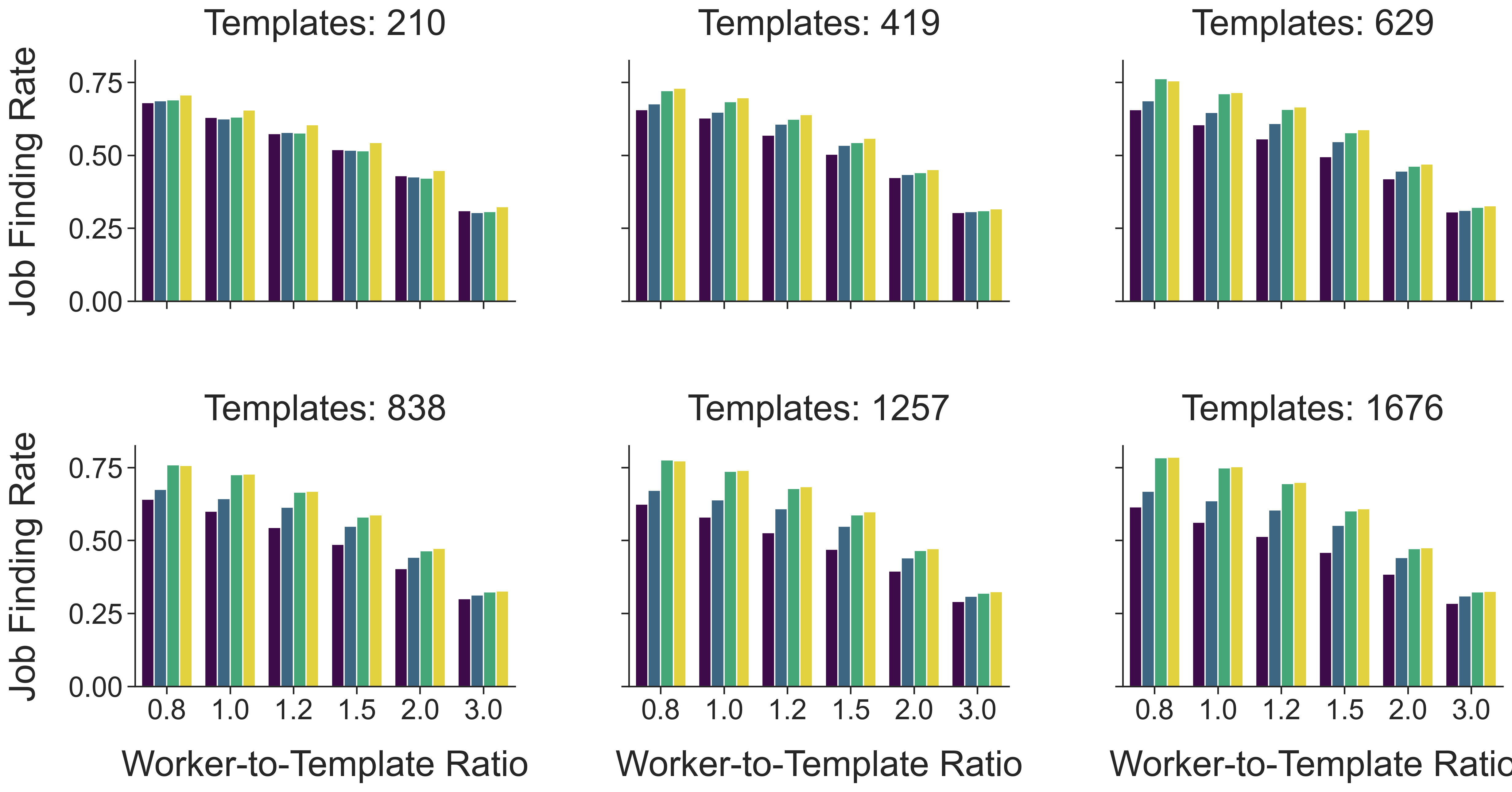}
    \includegraphics[width=0.45\textwidth]{figuretable/recommendation_cap_adjustment_project/ec_graph/legend_only.png}
    \caption{Per-Round Job-Finding Rates across Market Sizes and Worker-to-Template Ratios}
    \label{fig:simulation_robustness}
\end{figure}

Finally, we demonstrate that \AQ and \TEC perform well across a wide range of environments. Figure~\ref{fig:simulation_robustness} reports the average per-round job-finding rate under different market conditions, varying the balance between the number of workers and the number of templates. In the main simulation setting, the number of templates and workers is fixed at 838 and 1,000. In this robustness exercise, we consider markets with varying numbers of templates, set to 25\%, 50\%, 75\%, 100\%, 150\%, and 200\% of the baseline. For each case, the set of templates present in the market is constructed by sampling with replacement from the original 838 templates in the data. The resulting number of templates is shown at the top of each subfigure. For each template count, we vary the number of workers to be 80\%, 100\%, 120\%, 150\%, 200\%, and 300\% of the number of templates, as indicated on the horizontal axis within each subfigure. For each parameter combination, we simulate 10 different random seeds and plot the average per-round job-finding rate across these runs. Throughout this exercise, the parameter settings for \SQ, \AQ, and \TEC are held fixed.

Across the full range of market conditions in Figure~\ref{fig:simulation_robustness}, the adaptive methods dominate the benchmarks: \AQ and \TEC achieve the highest average per-round job-finding rates, followed by \SQ, with \Greedy performing worst. This result suggests that the misdirected concentration of exposure induced by \Greedy is highly detrimental to employment outcomes. The performance gaps also become more pronounced as the number of templates increases (e.g., in the bottom-right panel). When the market is small, the recommendation capacity $c = 16$ is substantial relative to the number of available templates, so even \Greedy gives many templates some exposure. In larger markets, by contrast, the same $c$ covers only a small fraction of the template space, and many templates receive little or no exposure under \Greedy, making misdirected concentration more harmful. Finally, across all configurations, \AQ and \TEC deliver nearly identical job-finding rates. Given that \TEC is fully parallelizable and much easier to implement in practice, these results suggest that \TEC is the more desirable approach for large-scale deployment.

The relative performance of \SQ versus \AQ/\TEC varies systematically across market conditions. The gap is comparatively small near the baseline configuration, where the number of templates is 838, and the number of workers is roughly 1.2 times as large, but it widens when the number of templates increases and/or when the worker-to-template ratio falls below 1.2. This pattern is expected because the fixed quota parameter in \SQ (i.e., $q = 25$) is tuned to the baseline setting and is not chosen to be optimal across all environments. Although \SQ still outperforms \Greedy in every configuration, matching the performance of \AQ or \TEC would require careful parameter tuning tailored to local market conditions. Such tuning is operationally difficult: the ratio of workers to templates varies over time, differs across regions, and does not admit clean geographic segmentation with stable ratios. Consequently, manually adapting parameters to each environment is infeasible in practice. These results underscore the practical value of adaptive exposure-allocation policies, such as \AQ\ and \TEC, which adjust automatically in response to changing market conditions.

\begin{remark}
    We also use the model to study the transition path. This short-run analysis is motivated by the fact that the results of the field experiment are observed over a finite calendar window and therefore need not reflect the new steady state. Because the transition results are most informative when interpreted jointly with the field-experimental evidence, we defer them until after presenting the experimental results in Section~\ref{sec:field_results_transition}. There, we use the transition analysis to clarify which margins should be expected to move within a one-month rollout and which effects would likely become visible only under a longer implementation horizon.
\end{remark}

\section{Field Experiment}\label{sec:field_experiment}

The simulation study identifies a central prediction to be tested in the field: improving spot-work recommendations requires reallocating exposure toward templates with unmet labor demand, rather than simply maximizing the predicted favoriting probabilities. In the simulated market, \TEC achieves this reallocation by controlling per-template exposure based on recent posting activity and unfilled capacity. This mechanism is especially relevant in Timee's environment, where favoriting is only an intermediate conversion outcome: actual matches are formed on an FCFS basis and are constrained by offering availability, worker schedules, and competition among subscribers.

These simulation results are informative, but they rely on predicted behaviors and abstract from actual responses that may arise in live platforms. In practice, recommendations could affect user behavior in various ways, and changes in exposure can generate spillovers across users on both sides of the market.

To assess whether the mechanisms identified in the simulation operate in a real-world environment, we conduct a prefecture-level randomized rollout of \TEC at production scale. The design is motivated by two practical considerations. First, recommendation policies on matching platforms can create interference among users; thus, treatment should be assigned at the prefecture level rather than at the user level. Second, because the rollout includes only one treated and one control prefecture, we use DID-style regressions mainly as convenient ITT estimators that absorb common time shocks and summarize post-period differences. Our goal is therefore to present causal evidence consistent with randomized assignment while keeping inference appropriately cautious.

\subsection{Experimental Setting}

\subsubsection{Experimental Setup}
We conducted the field experiment in two geographically segmented prefectures and randomly assigned treatment status at the prefecture level: Aomori was assigned to treatment, and Iwate to control. We chose a prefecture-level rollout instead of user-level randomization because recommendation changes can generate spillovers within a local market: treated users may redirect favorites and applications toward templates that control users would otherwise have seen, while templates' experiences may change as the composition of inbound favorites shifts. Prefecture-level assignment therefore limits within-market contamination relative to a fully randomized user-level experiment.\footnote{User-level randomization would create interference because treated and control users would compete within the same local market. A change in recommendations for treated users can shift their favoriting and application behavior toward different templates, thereby changing subscriber stocks, the speed at which offerings are filled, and the opportunities available to control users. In that case, control outcomes would no longer represent outcomes under the untreated recommendation policy. Such equilibrium and congestion effects are central in two-sided recommendation and matching settings \citep{rios2023improving,manshadi2023redesigning}. Prefecture-level assignment does not eliminate all possible interference, but it substantially reduces within-market contamination while remaining operationally feasible.}

The deployment started on Monday, January 12, 2026, and continued for a month through February 11, 2026. During this period, the default recommendation logic for users in Aomori was switched from the status quo, \Greedy, to our proposed \TEC. Formally, treatment was applied whenever the user's activity location (latitude and longitude captured at query time) fell within Aomori's administrative boundary. The control prefecture, Iwate, continued to operate under the status quo recommendation logic throughout the same period. Users were not informed about which recommendation logic was active in their prefecture.

The two prefectures are geographically adjacent. However, because Aomori and Iwate are rural prefectures separated by mountainous terrain and limited transportation infrastructure, cross-border commuting is negligible, making them effectively disjoint as spot-work labor markets. At the same time, they share similar job compositions in terms of industry mix, wage levels, and posting patterns on the platform, which makes them a natural treatment--control pair.

For the pre-treatment comparison window, we use January 2, 2026 through January 11, 2026. This choice reflects the platform's operational knowledge that the period from Christmas through the end of December is affected by year-end demand surges, while the New Year holiday is unusually slack. January 2--11 is regarded as the first relatively stable post-holiday period before deployment. Restricting the pre-period to this window, therefore, helps avoid comparing the rollout against atypical holiday fluctuations in worker activity and labor demand.

Based on the simulation results, the treatment adopted \TEC with the parameter settings used in the simulation: average quota $\bar{q} = 40$, score weight for posting $w^0 = 50$, and score weight for unfilled capacity $w^1 = 125$. As discussed in Section~\ref{sec:simulation}, these parameters were the best for the baseline simulation setting among those tested, and \TEC adaptively responded to market conditions without manual retuning, based on recent posting activity and unfilled labor demand. As a result, the introduction of \TEC was considered to have a low operational risk from a business perspective.

\subsubsection{Outcome}\label{subsubsec:experiment_outcome}
We evaluate the rollout using outcomes that track the recommendation-to-match pipeline: exposure to templates, favorite-list formation, and applications to offerings. Because treatment is assigned at the prefecture level and the deployment data are recorded by calendar day, our main aggregate outcomes are measured at the prefecture-day level, which provides the natural field counterpart to the round-level outcomes in the simulation. We also use impression-level data to examine how individual recommendation exposures translate into favoriting and subsequent matching behavior.

First, we consider prefecture-day level outcomes, which are aggregated from template-level and offering-level data and quantify market-level engagement and fulfillment. To link with the simulation in Section~\ref{sec:simulation} explicitly, the average number of recommended workers per template is the field counterpart of recommendation flow per template, and the average number of subscribers per template is the field counterpart of the subscriber stock generated by favoriting behavior. We also define \emph{active-offering fulfillment} in prefecture $p$ on day $t$ as total matched slots divided by total posted capacity across active offerings in that prefecture-day. This is the closest empirical analogue available in the field data of the per-round fill rate in the simulation.\footnote{Because we do not observe whether a worker searched for offerings for a given day, it is difficult to construct a direct empirical analogue of the per-round job-finding rate in the simulation.} In addition, we report the total numbers of favorites and matches per prefecture-day as market-level outcomes that summarize how the rollout affected aggregate engagement and realized employment.

Second, we examine impression-level outcomes, which are based on individual recommendation exposures on the screen, introduced in Section~\ref{sec:user_interface}. For each on-screen recommendation (impression), we define two binary outcomes: (i) whether the impression led the user to add the template to the favorite list, and (ii) conditional on being favorited, whether that favorited impression subsequently resulted in a match. The second outcome is therefore estimated by comparing favorited impressions observed in the pre-treatment period with favorited impressions observed in the post-treatment period across treatment and control prefectures. Unlike prefecture-day aggregates, these measures capture how changes in exposure translate into observed favoriting and matching behavior at the unit of each recommendation impression and can only be obtained through a live deployment.

\subsubsection{Balance Table of Workers and Offerings}
Table~\ref{tb:balance_table} reports summary statistics for workers and offerings in the two experimental prefectures during the study period. Panel~(a) shows that the worker pools are similar in size and demographic composition: mean age is virtually identical at approximately 42 years in both prefectures, although Aomori has a somewhat higher female ratio (0.60 versus 0.53). Panel~(b) compares offering-level characteristics. Mean hourly wages are comparable (1,074 JPY in Iwate versus 1,049 JPY in Aomori), though the wage distribution in Iwate exhibits considerably more dispersion. Average working hours per offering are slightly longer in Iwate (4.63 versus 4.30 hours), and transportation reimbursements are higher in Iwate on average (293 versus 166 JPY). These differences are modest relative to the within-prefecture variation. Because the treatment was randomized at the prefecture level with only two units, observable balance cannot be guaranteed by design. The similarity in worker demographics and offering characteristics nonetheless provides reassurance that the two prefectures constitute comparable spot-work markets, reducing concerns that differences in market composition, rather than the algorithm change, drive the estimated treatment effects.

\begin{table}[t!]
  \begin{center}
      \caption{Balance of Worker and Offering Characteristics}
      \label{tb:balance_table}
      \subfloat[Workers]{
\begin{tabular}{lrr}
\toprule
 & Iwate (Control) & Aomori (Treatment)\\
\midrule
Age & 42.21 & 42.25\\
 & (12.92) & (13.00)\\
Female ratio & 0.53 & 0.60\\
\bottomrule
\end{tabular}
}\\
      \bigskip
      \subfloat[Offerings]{
\begin{tabular}{lrr}
\toprule
 & Iwate (Control) & Aomori (Treatment)\\
\midrule
Hourly wage (JPY) & 1073.75 & 1049.31\\
 & (336.71) & (104.63)\\
Working hours & 4.63 & 4.30\\
 & (1.89) & (1.74)\\
Transport reimbursement (JPY) & 293.22 & 165.63\\
 & (255.70) & (210.94)\\
\bottomrule
\end{tabular}
}
  \end{center}\footnotesize
  \textit{Note}: Mean (SD) reported. Workers are active users in each prefecture during the experiment period (January 2--February 11, 2026). Offerings are job slots posted in each prefecture during the same period, aggregated at the day level. Hourly wage and transportation reimbursement are in Japanese yen (JPY). Exact sample sizes are confidential and cannot be reported, but the number of unique users is on the order of a few thousand, and the number of unique offerings is just under ten thousand in each prefecture, with treatment and control groups confirmed to be comparable in scale.
\end{table}

\subsection{Methodology}

\subsubsection{Prefecture-Day-Level Analysis: Difference-in-Differences Style Specification}

With treatment status assigned at the prefecture level, we estimate the assignment-based ITT effects using a DID-style specification that compares before-after changes in the treated prefecture with those in the control prefecture. This regression is useful because it absorbs common day shocks and provides a compact summary of post-period treatment effects, but the identifying variation comes from the randomized assignment of treatment status rather than from a conventional many-region parallel-trends argument. With a slight abuse of notation, let $p\in\{\text{AOM},\text{IWA}\}$ index prefectures and $t$ index calendar days. For an outcome $Y_{tp}$ observed at the prefecture-day level, our baseline specification is
\begin{equation}
\label{eq:did-baseline}
Y_{tp}
= \alpha_p + \delta_t
+ \beta_{\text{ITT}}\;\mathbf{1}\{p=\text{AOM}\}\times\mathbf{1}\{t\in\mathcal{T}_{\text{deploy}}\}
+ \varepsilon_{tp},
\end{equation}
where $\alpha_p$ are prefecture fixed effects and $\delta_t$ are day fixed effects capturing common shocks (seasonality, holidays, marketing, etc.), and $\mathcal{T}_{\text{deploy}}$ is the set of treatment days.


\paragraph{Identification Assumptions}
Identification of $\beta_{\text{ITT}}$ in Equation~\eqref{eq:did-baseline} relies primarily on the randomized assignment of treatment status across the two rollout prefectures. In this sense, the design is best viewed as a cluster-randomized rollout analyzed with a DID estimator. The key assumptions are therefore: (i) \textit{randomized implementation}: treatment status was assigned as planned and the recommendation logic switched only in the treated prefecture during the deployment window; (ii) \textit{no anticipation}: outcomes did not respond before the first deployment day $t_0$; (iii) \textit{stable composition and measurement}: the intervention did not differentially change how outcomes were recorded or which users/templates were in scope across prefectures; and (iv) \textit{limited interference}: spillovers between Aomori and Iwate were small. These assumptions are supported by the app-based prefecture-level rollout design described in Section~\ref{sec:design_discussion}.

\subsubsection{Impression-Level Analysis: Linear Probability Model}

At the impression level, we treat each recommendation exposure on the screen as the unit of analysis and evaluate how the treatment affects user responses by linking impression logs with favorite and match records. We estimate linear probability models that control for recommendation ranked position, date, and prefecture, and interpret the treatment--control contrast as an impression-level ITT effect induced by the randomized prefecture rollout, holding displayed ranked position fixed.

\subsubsection{Template-Day-Level Analysis: Distribution-Regression DID}

The mean DID specification in Equation~\eqref{eq:did-baseline} summarizes first-moment effects but does not show where in the template-day-level distribution the treatment operates. To study that margin, we estimate a distribution-regression DID following \cite{fernandez2024distribution}. Let $Y_{kpt}$ denote either recommendations per template or subscribers per template for template $k$ in prefecture $p$ on day $t$, and let $Y_{kpt}(0)$ and $Y_{kpt}(1)$ denote the corresponding potential outcomes under the untreated and treated recommendation policy, respectively. The observed outcome satisfies $Y_{kpt}=Y_{kpt}(0)(1-D_p P_t)+Y_{kpt}(1)D_p P_t$, where $D_p=\mathbf{1}\{p=\text{AOM}\}$ is the prefecture treatment indicator and $P_t=\mathbf{1}\{t\in\mathcal{T}_{\text{deploy}}\}$ is the post-period indicator. For each threshold $y$, we estimate the saturated logit model
\begin{equation}\label{eq:drdid}
  \Pr(Y_{kpt}\le y \mid D_p,P_t)
  =
  \Lambda\!\left(
  \alpha(y)
  + \beta(y) P_t
  + \gamma(y) D_p
  + \theta(y) D_p P_t
  \right),
\end{equation}
where $\Lambda(\cdot)$ is the logistic CDF.

\paragraph{CDF Notation}
Adopting the notation of \cite{fernandez2024distribution}, we write
\begin{equation}
    F_{Y(d)\mid g,t}(y) := \Pr(Y_{kpt}(d) \le y \mid D_p=g, P_t=t)
\end{equation}
for the conditional CDF of potential outcome $Y_{kpt}(d)$ for $d \in \{0,1\}$, $g \in \{0,1\}$, $t \in \{0,1\}$, and abbreviate the treated-post cell ($g=t=1$) as ``$\mid 11$''. Thus $F_{Y(1)\mid 11}(y)$ is the CDF of the realized outcome under treatment in the treated prefecture during the post period, while $F_{Y(0)\mid 11}(y)$ is the (untreated) counterfactual CDF for the same group-period cell. The distribution-regression DID restriction is imposed on the untreated potential-outcome surface: absent treatment, the group and post-period effects in the log-odds CDF are additive at each threshold $y$ \citep[][Assumption~1]{fernandez2024distribution}. Under this restriction, these two CDFs are identified by
\begin{align}
    \widehat F_{Y(1)\mid 11}(y) &= \widehat\Pr(Y_{kpt}\le y \mid D_p=1,P_t=1),\\
    \widehat F_{Y(0)\mid 11}(y) &= \Lambda(\hat{\alpha}(y)+\hat{\beta}(y)+\hat{\gamma}(y)).
\end{align}
We estimate the distributional treatment effect at each threshold $y$ as
\begin{equation}
    \widehat{\Delta}_F(y)=\widehat F_{Y(1)\mid 11}(y)-\widehat F_{Y(0)\mid 11}(y).
\end{equation}
A negative $\widehat{\Delta}_F(y)$ means that treatment reduces the share of template-days at or below threshold $y$, shifting probability mass toward larger outcomes. We report pointwise 95\% confidence intervals at each threshold using the weighted bootstrap procedure of \cite{fernandez2024distribution} (Algorithm~3) with 500 replications.



\subsection{Experimental Design Discussion}\label{sec:design_discussion}

The remaining design question is why treatment was assigned at the prefecture level rather than at a finer geographic unit. Because the intervention changes exposure to templates, finer geographic assignment could create boundary contamination. Users may search for jobs near administrative borders, and postings located near a boundary may be shown under different recommendation rules depending on the user's query location. Such contamination would blur the distinction between treated and control markets even if treatment were assigned geographically.

We therefore used prefectures as the geographic unit of rollout. This choice substantially reduces within-market interference while keeping the treatment definition transparent and implementable. In addition, platform internal pre-experiment data indicated that activity along the Aomori--Iwate border was limited. We therefore expect spillovers generated by cross-border search or application behavior to be small.

This design also has an important limitation. Because the rollout includes only one treated and one control prefecture, the estimates are best interpreted as evidence from this realized rollout rather than as estimates supported by large-cluster asymptotics. This limitation motivates our use of DID-style specifications as compact post-period summaries rather than as a basis for strong asymptotic inference.

\subsection{Results}

\subsubsection{Prefecture-Day-Level Results}\label{sec:prefecture_day_results}

Table~\ref{tb:prefecture_day_did_combined} reports ITT estimates from the prefecture-day-level DID regressions. The treatment-prefecture and post-period main effects are absorbed by prefecture and date fixed effects and are not reported.

\begin{table}[!tp]
  \begin{center}
      \caption{Difference-in-Differences Estimates}
      \label{tb:prefecture_day_did_combined}
      \subfloat[Prefecture-Day Level (Fav Count, Apply Count)]{
\begin{tabular}[t]{lcc}
\toprule
  & Fav & Match\\
\midrule
Treatment $\times$ After & -40.458 & 9.045**\\
 & (51.977) & (4.374)\\
\midrule
Num.Obs. & 82 & 82\\
R2 & 0.693 & 0.849\\
R2 Adj. & 0.362 & 0.687\\
FE (Date) & Y & Y\\
\bottomrule
\end{tabular}
}\\
      \bigskip
      \subfloat[Prefecture-Day Level (Avg Rec Workers, Avg Subscribers, Per-Round Fill Rate)]{
\begin{tabular}[t]{lccc}
\toprule
  & \makecell{Avg Rec Workers\\(Active)} & \makecell{Avg Subscribers\\(Active)} & \makecell{Per-Round Fill Rate\\(Active)}\\
\midrule
Treatment $\times$ After & 0.571** & -0.082 & -0.002\\
 & (0.253) & (0.056) & (0.007)\\
\midrule
Num.Obs. & 82 & 82 & 82\\
R2 & 0.760 & 0.661 & 0.926\\
R2 Adj. & 0.502 & 0.295 & 0.847\\
FE (Date) & Y & Y & Y\\
\bottomrule
\end{tabular}
}
  \end{center}\footnotesize
  \textit{Note}: Panels (a) and (b) report prefecture-day level regressions with prefecture and date fixed effects; only the treatment-by-post interaction is reported. The coefficients are interpreted as ITT effects from a randomized prefecture-level rollout, estimated with DID-style regressions. In panel (b), \emph{(Active)} indicates that the outcome is computed over active templates (those with at least one impression on that day) for the first two columns, or over active offerings (those opened on that day) for the third column. \emph{Avg Rec Workers (Active)} is the prefecture-day mean of the number of distinct workers recommended a given active template. \emph{Avg Subscribers (Active)} is the prefecture-day mean of the number of distinct workers who favorited a given active template via the recommender. \emph{Per-Round Fill Rate (Active)} is the prefecture-day ratio of total matched slots (workers with work status applied or completed) to total posted capacity across active offerings.\\ Standard errors in parentheses are heteroskedasticity- and autocorrelation-consistent, allowing for serial correlation in daily prefecture-level outcomes. * $p<0.10$, ** $p<0.05$, *** $p<0.01$.
\end{table}

Panel~(a) shows that the rollout did not generate a statistically precise change in the daily number of favorites in the prefecture, although the treatment $\times$ after coefficient for favorites is $-40.458$. This absence of a detectable aggregate effect on favorites is not a failure of the recommender design. \TEC was not introduced to mechanically raise favoriting counts, so a null effect on this margin is consistent with the algorithm's objective and also provides some reassurance that the rollout did not materially reduce favoriting activity at the market level. By contrast, the same interaction term is positive and statistically significant for matches, with a DID estimate of $9.045$ matches per prefecture-day.\footnote{The corresponding baseline number of matches is not reported due to confidentiality restrictions.} Note also that the transition-path simulation introduced in Section~\ref{sec:field_results_transition} suggests that a longer rollout period would plausibly allow the cumulative effect to grow as the reallocation of exposure gradually feeds through to subscriber stocks and downstream matching.

Panel~(b) turns to the template-level outcomes that were also evaluated in the simulation study with analogous variables. As discussed in Section~\ref{sec:simulation}, the simulation analysis tracked how alternative recommenders changed recommendation exposure and subscriber stocks across templates. The corresponding field estimates show that \TEC increased the average number of recommended workers per template by $0.571$, while the estimates for average subscribers per template and per-round fill rate are small and statistically insignificant.

An important measurement note is that the statistics are computed only over templates that appear in the impression log on a given day---that is, templates that received at least one recommendation. Templates that were not recommended to any user are excluded from this average. The positive coefficient therefore reflects \TEC concentrating recommendation exposure on \emph{active} templates with greater labor demand, which is the algorithm's intended behavior: by redirecting exposure toward offerings that are actually posted and seeking workers, \TEC raises the per-template mean among the set of templates that the algorithm chose to feature.

These results are consistent with the timing implied by the recommendation-to-match mechanism. The most immediate effect of \TEC is on recommendation exposure: the algorithm directly changes which templates are eligible to be shown and how exposure is allocated across templates. The positive estimate for active recommendations therefore provides the clearest short-run evidence that the field rollout operated on the intended margin. Subscriber counts, by contrast, are a stock variable that changes only through the accumulation of exposure and subsequent favoriting decisions. It is therefore not surprising that the mean number of subscribers per template does not move detectably within the relatively short experimental window. This mean result does not imply that subscriber formation was unaffected; rather, any early changes are more likely to appear first at the extensive margin of the distribution, which we examine in Section~\ref{sec:distributional_results}, and to accumulate more clearly over a longer horizon, as studied in Section~\ref{sec:field_results_transition}.

We also examine the active-offering per-round fill rate, defined as total matched slots (workers with work status ``applied'' or ``completed'') divided by total posted capacity across active offerings in a given prefecture-day. The DID estimate is $-0.002$ and statistically indistinguishable from zero. This null result is also consistent with the same timing logic. Fulfillment is further downstream than subscriber formation: recommendations must first alter exposure, exposure must then accumulate into subscriber stocks, and those subscribers must subsequently observe, choose, and apply to available offerings. A one-month rollout may therefore be long enough to reveal changes in recommendation exposure, and even some realized matches, but too short for the induced change in favorite-list stocks to fully propagate into active-offering fulfillment. Note also that the baseline per-round fill rate in both experimental prefectures was already high---approximately 84--85\% during the pre-period---leaving limited room for short-run improvement on this margin. In contrast, the simulation study documents large fulfillment gains in an environment with a lower baseline per-round fill rate (67.4\% under \Greedy). Taken together, the field and simulation results suggest that \TEC's fulfillment effects are likely to be more visible under a longer implementation horizon, and especially in markets where baseline fulfillment is lower and misdirected concentration under \Greedy leaves greater scope for exposure reallocation to improve market-level employment outcomes.

\subsubsection{Impression-Level Results}

Table~\ref{tb:impression_did_combined} reports impression-level linear probability models. Column~(1) shows that the probability that an impression leads to a favorite rises by $0.005$ under \TEC, controlling for recommendation ranked position, date, and prefecture fixed effects. This increase in favoriting probability is a favorable and somewhat unexpected finding. \TEC is designed to reallocate exposure toward active templates with unmet labor demand, rather than to maximize the probability that a displayed template is favorited. Indeed, because the status quo \Greedy recommender directly targets predicted favoriting probabilities, one might have expected \Greedy to perform better on this intermediate conversion margin. The positive impression-level effect therefore suggests that \TEC not only redirected exposure toward templates with greater labor demand, but also improved the quality of the recommendation set faced by workers in a way that increased their propensity to favorite displayed templates.

\begin{table}[tp!]
  \begin{center}
      \caption{Impression-Level Analysis}
      \label{tb:impression_did_combined}
      
\begin{tabular}[t]{lcccc}
\toprule
  & Fav & Fav  & Fav   & Match via Fav\\
\midrule
Treatment $\times$ After & 0.005*** & 0.005*** & 0.008*** & 0.001**\\
 & (0.001) & (0.001) & (0.001) & (0.001)\\
1(Open Slot) &  & -0.008*** & -0.008*** & 0.000\\
 &  & (0.001) & (0.001) & (0.001)\\
Repeat Views &  &  & -0.011*** & \\
 &  &  & (0.001) \vphantom{3} & \\
Treatment $\times$ Repeat Views &  &  & 0.001 & \\
 &  &  & (0.001) \vphantom{2} & \\
After $\times$ Repeat Views &  &  & 0.009*** & \\
 &  &  & (0.001) \vphantom{1} & \\
\makecell[l]{Treatment $\times$ After\\$\times$ Repeat Views} &  &  & -0.002* & \\
 &  &  & (0.001) & \\
\midrule
Num.Obs. & 814432 & 814432 & 814432 & 298184\\
R2 & 0.011 & 0.011 & 0.012 & 0.003\\
R2 Adj. & 0.011 & 0.011 & 0.012 & 0.003\\
Recommend Rank & Y & Y & Y & Y\\
FE (Date) & Y & Y & Y & Y\\
FE (Prefecture) & Y & Y & Y & Y\\
\bottomrule
\end{tabular}

  \end{center}\footnotesize
  \textit{Note}: The table reports impression-level linear probability models with recommendation rank, date, and prefecture fixed effects; treatment-prefecture and post-period main effects are not reported. The match specification is estimated on the sample of favorited impressions only. \\ Standard errors in parentheses. * $p<0.10$, ** $p<0.05$, *** $p<0.01$.
\end{table}

A natural first hypothesis is that this improvement arises because workers value templates with currently available offerings, as shown in Figure~\ref{fig:app_screenshot}. If workers use the recommendation tab partly to find templates that are likely to generate work opportunities, then templates with open slots may be more attractive, and \TEC tends to recommend such templates more often. To examine this hypothesis, in Column~(2), we control for whether the recommended template had at least one open slot on the day of the impression. However, the estimate does not support this interpretation. The coefficient on $\mathbf{1}(\text{Open Slot})$ is significantly negative ($-0.008$), meaning that, if anything, templates with visible open capacity are \emph{less} likely to be favorited. Furthermore, controlling for slot availability does not absorb the treatment effect---the treatment $\times$ after coefficient remains positive and significant at $+0.005$. Thus, the increase in favoriting probability under \TEC does not appear to operate through a simple ``vacancy visibility'' channel. Instead, the result suggests that the treatment improved recommendation quality through another mechanism.

This negative association between visible open capacity and favoriting also provides a useful perspective on how workers interpret capacity-related information. Prior work on availability disclosure has often emphasized that low availability can stimulate demand through scarcity signaling \citep{calvo2023disclosing,knight2026disclosing}, while related work in labor markets shows that job seekers may \emph{avoid} postings with high observed competition \citep{fradkin2026competition}. Our findings point to a distinct margin in capacity-constrained recommendation environments: visible remaining capacity may be interpreted not only as an opportunity to act, but also as a signal that the template is less attractive or less popular. Thus, simply surfacing available capacity need not increase engagement. In our setting, workers appear less likely to favorite templates with open slots, suggesting that the positive favoriting effect of \TEC comes from changes in the composition and novelty of recommendations rather than from workers responding mechanically to current availability.

A second hypothesis is that the improvement comes from reducing repetition in workers' recommendation lists. Because \Greedy is optimized for predicted favoriting probabilities, it may repeatedly show workers the same templates that look attractive according to the prediction model. By contrast, \TEC adjusts exposure priorities in response to market conditions, which may refresh the set of templates shown to workers and increase the value of each impression. Column~(3) examines this mechanism by adding a triple interaction between treatment, after, and the number of previous days the user has already seen the same template (\emph{Repeat Views}). Repeat viewing strongly reduces favoriting ($-0.011$ per additional day), as expected: templates that a worker has already seen and declined to favorite offer diminishing novelty. The treatment $\times$ after coefficient in this specification is $0.008$, which is the estimated treatment effect at \emph{Repeat Views} $=0$, while the negative triple interaction ($-0.002$) indicates that the treatment-control difference is smaller for repeated exposures. Together with lower average repeat views in treated post-period impressions ($1.19$ versus $1.35$ in control), these patterns are consistent with \TEC presenting workers with a broader, less repetitive set of templates. They do not by themselves prove a mediation channel, but they suggest that novelty is one margin through which \TEC may raise the per-impression favoriting probability.

Finally, Column~(4) reports the match outcome, estimated on the sample of favorited impressions only. The sample is therefore much smaller than in Columns~(1)--(3)---restricted to the subset of impressions that were favorited rather than the full set of impressions---and the dependent variable now measures, conditional on favoriting, whether the favorited template subsequently led to a successful match. The treatment $\times$ after coefficient is $0.001$ and statistically significant, indicating that conditional on favoriting, treated impressions are more likely to lead to downstream matches. This is consistent with \TEC's design objective of channeling exposure toward templates with unmet labor demand. Combined with the favoriting result, the impression-level evidence indicates that \TEC improved recommendation quality along two margins: workers were more likely to favorite the recommended templates, and, having favorited, more likely to secure matches from them.

\subsubsection{Distributional Results}\label{sec:distributional_results}

\begin{figure}[!tp]
  \begin{center}
      \begin{subfigure}{0.7\textwidth}
        \centering
        \includegraphics[width=\textwidth]{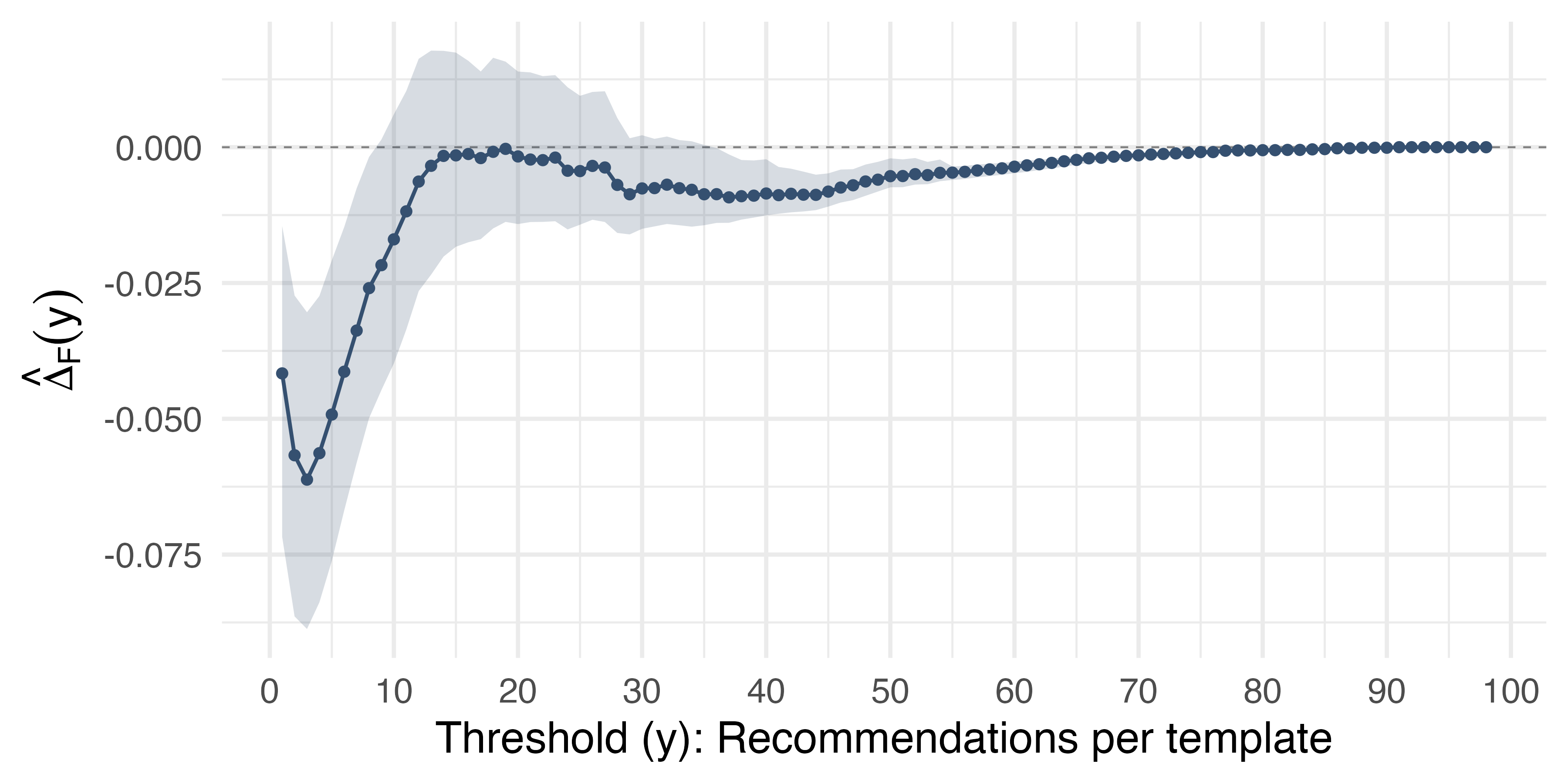}
        \caption{Recommendations per Template}
      \end{subfigure}\\
      \bigskip
      \begin{subfigure}{0.7\textwidth}
        \centering
        \includegraphics[width=\textwidth]{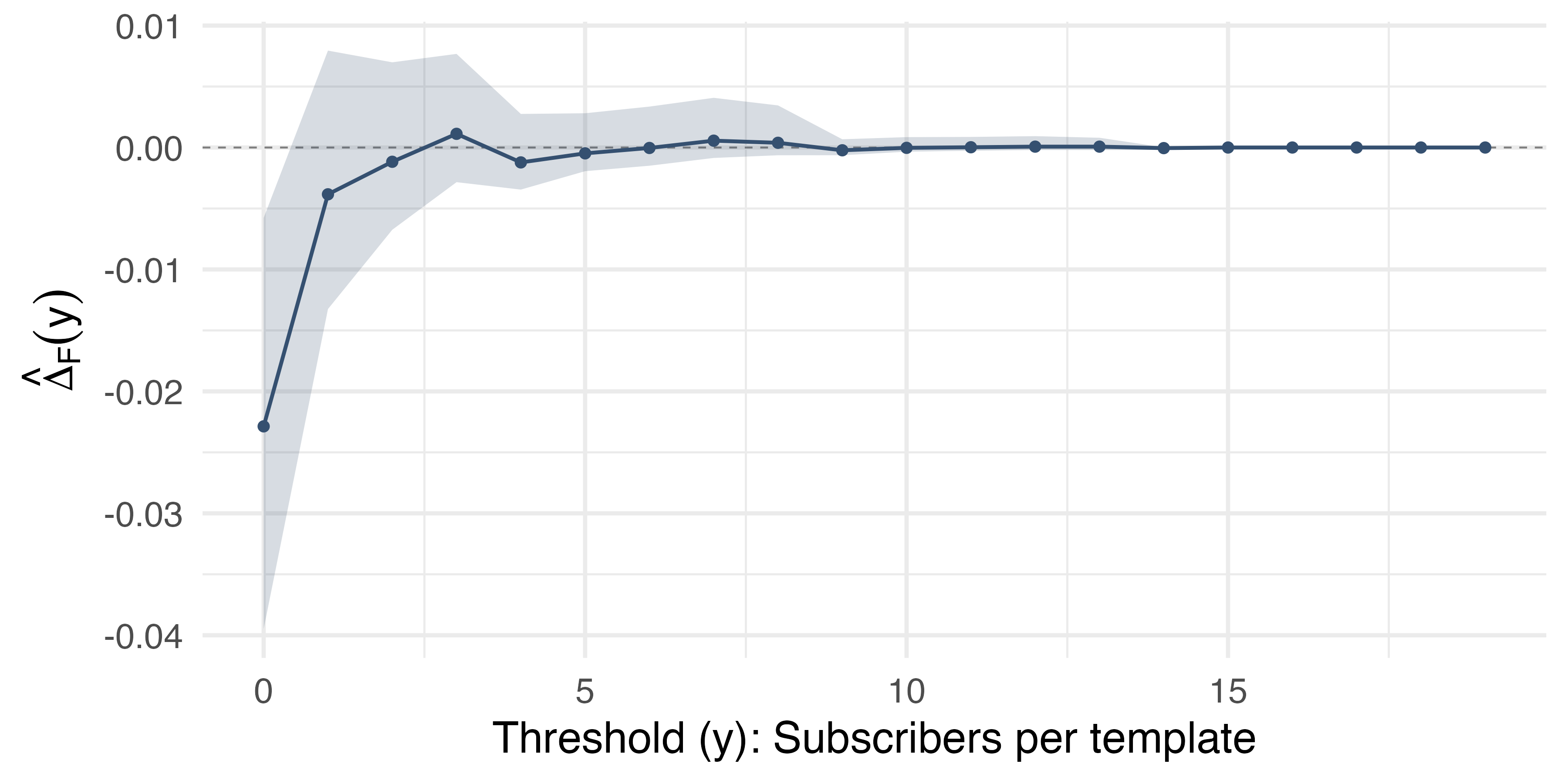}
        \caption{Subscribers per Template}
      \end{subfigure}
  \end{center}
  \caption{Distribution-Regression DID Treatment Effects}\footnotesize
  \textit{Note}: Each point plots the estimated distributional treatment effect $\widehat{\Delta}_F(y)$ from Equation~\eqref{eq:drdid} at threshold $y$. Shaded bands are pointwise 95\% confidence intervals based on 500 weighted bootstrap replications following \cite{fernandez2024distribution} (Algorithm~3). A negative value indicates that the treatment shifts probability mass above threshold $y$.
  \label{fig:drdid_delta_f}
\end{figure}

Figure~\ref{fig:drdid_delta_f} complements the mean DID estimates by showing where in the template-day distribution the treatment effect appears. For each threshold $y$, the figure plots $\widehat{\Delta}_F(y)$, the estimated difference between the observed treated-post CDF and the recovered untreated counterfactual CDF at that threshold. Thus, a negative value means that, relative to the counterfactual, fewer treated template-days lie at or below $y$; equivalently, probability mass has shifted toward larger values.

Panel~(a) shows a clear upward shift in the distribution of recommendations per template under \TEC. The estimated effect is most negative at $y=3$, where $\widehat{\Delta}_F(3)=-0.061$ with a 95\% confidence interval of $[-0.089,-0.030]$. This means that the rollout reduces the share of template-days with three or fewer recommendations by about six percentage points relative to the recovered counterfactual. This result is consistent with the positive mean effect on recommendations per template in Table~\ref{tb:prefecture_day_did_combined}: \TEC raises average exposure among active templates by moving mass out of the lower tail rather than only stretching the extreme right tail.

Panel~(b) reports the distributional effect for subscribers per template. The estimated CDF effect at zero is negative, indicating that, relative to the recovered counterfactual, the zero-subscriber mass of the template-day distribution is lower under \TEC. At thresholds above zero, the bootstrap intervals include zero. This null pattern at positive subscriber counts matches the mean specification in Table~\ref{tb:prefecture_day_did_combined}. 
This result is consistent with the experimental environment: in a relatively low-density rural market, where each template has access to a smaller nearby worker population, increasing subscriber counts beyond the first few workers may require a longer exposure period than the one-month rollout.

\subsection{Longer-Run Implications from Transition Simulation}\label{sec:field_results_transition}

The field experiment results (Table~\ref{tb:prefecture_day_did_combined}) show that the rollout of \TEC increased realized matches and the average number of recommended workers per template, while leaving daily favorites, average subscribers per template, and active-offering fulfillment without detectable movement at the prefecture-day level. At the same time, Table~\ref{tb:impression_did_combined} shows that treated recommendation impressions were more likely to generate both favorites and downstream matches. To interpret which of these patterns should be viewed as early transition effects and which would likely change under a longer implementation horizon, we return to the simulation model and study the transition path from \Greedy to \TEC. The analysis also sheds light on the importance of cumulative effects on long-run favorite-list management.

The simulation starts from the \Greedy steady state. We first run the economy under \Greedy until it reaches the steady state defined in Section~\ref{sec:simulation}, continue for 20 additional rounds to stabilize the pre-switch state, and then compare two counterfactual paths that share the same realized state at round 0. In one path, the platform continues to use \Greedy. In the other, it switches to \TEC. At the time of the switch, each template is assigned an initial score of 40, and \TEC thereafter uses the same parameter values as in the steady-state analysis, namely, $w^0=50$ and $w^1=125$. For each policy path, we generate 200 sample paths and simulate 500 post-switch rounds; the figures display the first 200 rounds, where the main adjustment occurs. Because the field experiment estimates average effects over a finite rollout window, we report cumulative means: for each outcome and horizon $t$, the plotted value is the average from round $0$ through round $t$. In each panel, the solid line plots the average, and the shaded areas indicate $\pm 2$ standard deviations across simulated sample paths.

\begin{figure}[!tp]
  \begin{center}
  \begin{subfigure}[t]{0.48\textwidth}
    \centering
    \includegraphics[width=\textwidth]{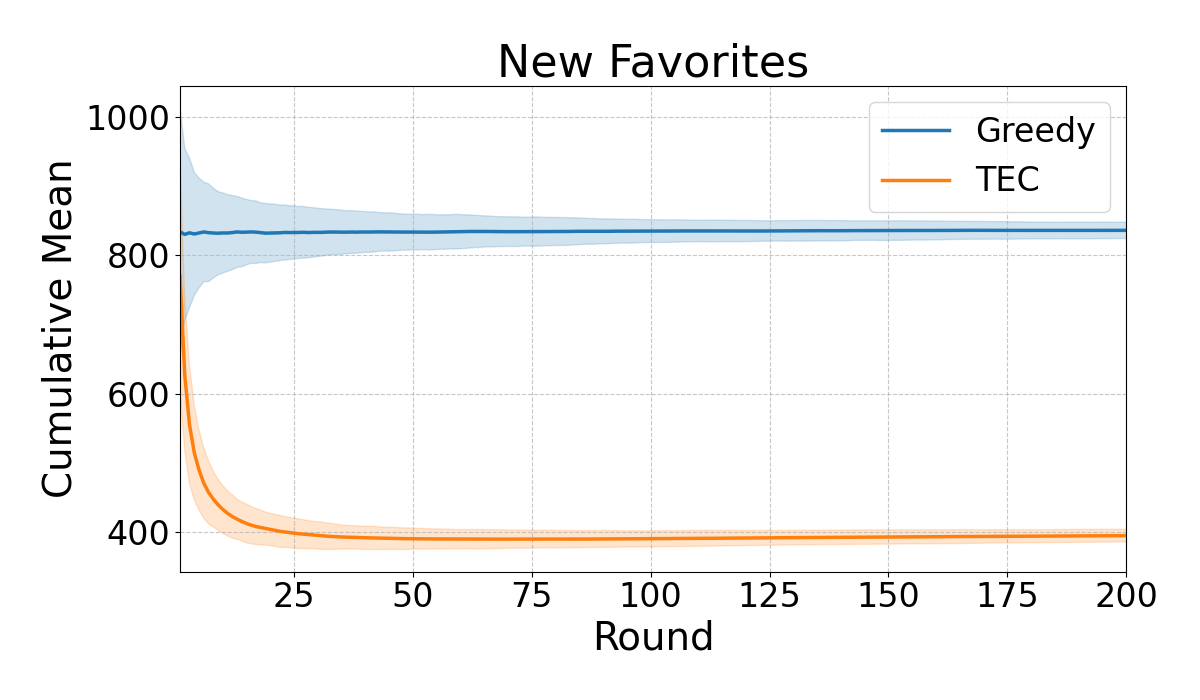}
    \caption{New Favorites}
    \label{fig:transition_new_favorites}
  \end{subfigure}
  \hfill
  \begin{subfigure}[t]{0.48\textwidth}
    \centering
    \includegraphics[width=\textwidth]{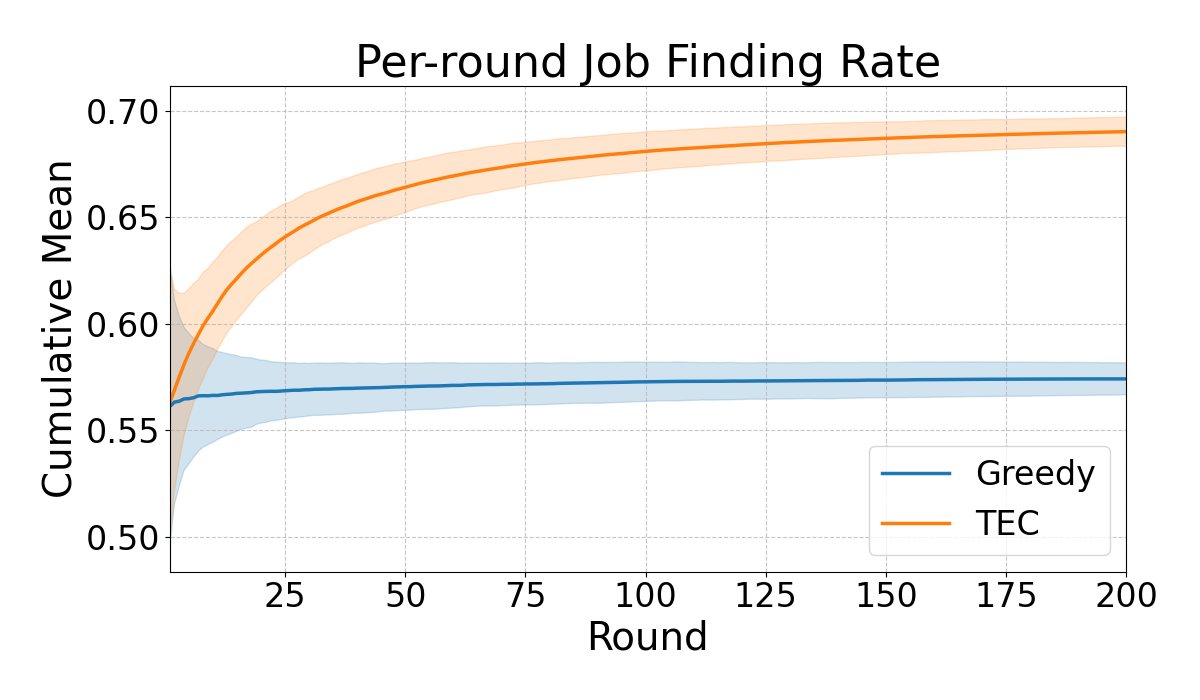}
    \caption{Per-Round Job-Finding Rate}
    \label{fig:transition_worker_activation}
  \end{subfigure}\\
  \begin{subfigure}[t]{0.48\textwidth}
    \centering
    \includegraphics[width=\textwidth]{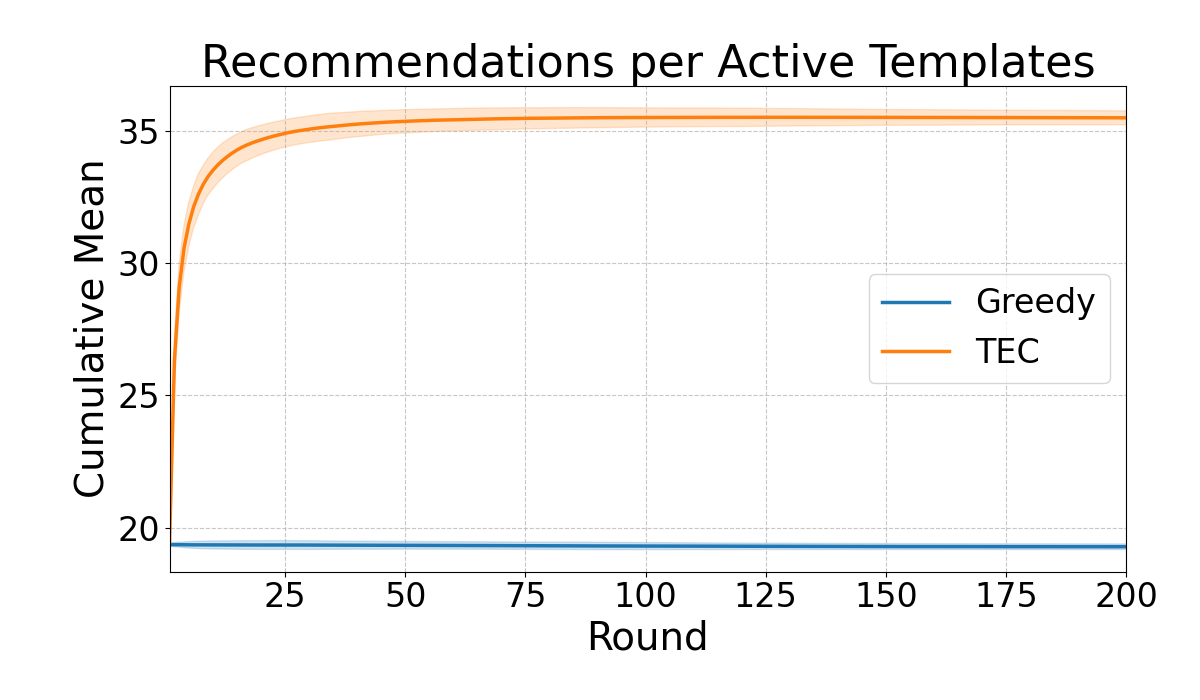}
    \caption{Recommendations per Active Template}
    \label{fig:transition_recommendations_per_active_template}
  \end{subfigure}
  \hfill
  \begin{subfigure}[t]{0.48\textwidth}
    \centering
    \includegraphics[width=\textwidth]{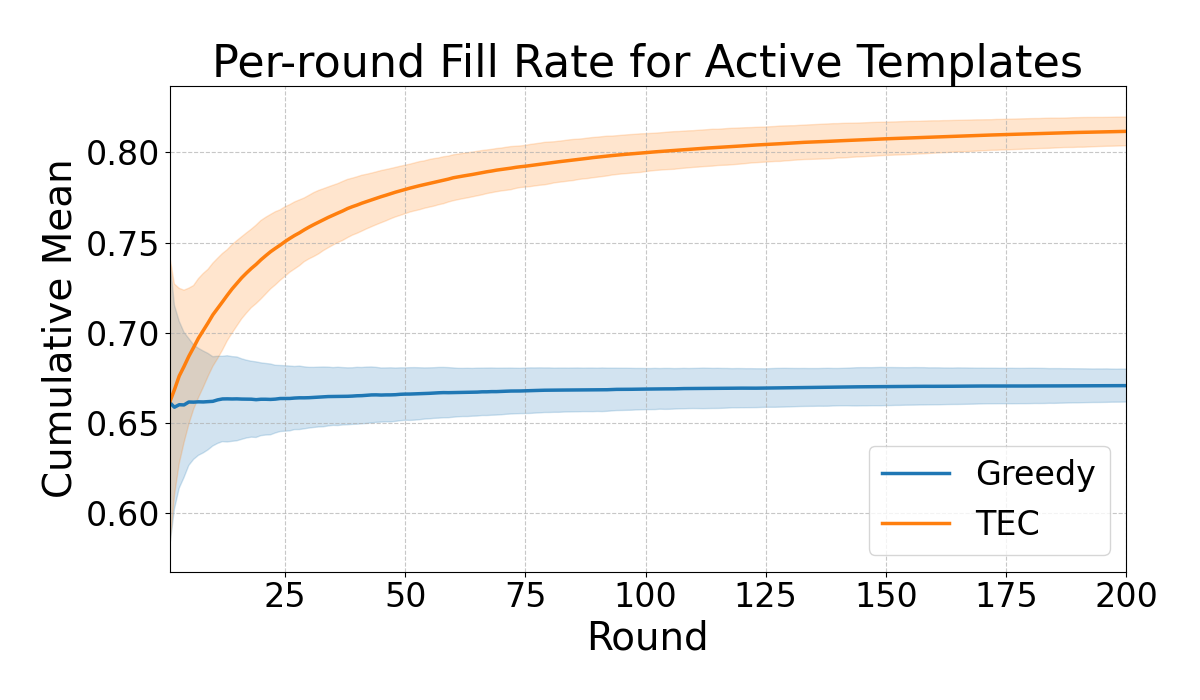}
    \caption{Per-Round Fill Rate for Active Templates}
    \label{fig:transition_fulfillment_per_active_template}
  \end{subfigure}
  \end{center}
  \caption{Transition from \Greedy to \TEC in the Simulation}\footnotesize
  \textit{Note}: Transition simulations from the \Greedy steady state. At round 0, the platform either continues with \Greedy or switches to \TEC. Each panel plots the cumulative mean of the indicated outcome from round 0 through horizon $t$, averaged over 200 simulated sample paths. Shaded areas indicate $\pm 2$ standard deviations across simulated sample paths.
  \label{fig:transition_results}
\end{figure}

\paragraph{New Favorites}
We first consider favorites. At the prefecture-day level, the field rollout did not produce a statistically significant change in the number of new favorites. However, Figure~\ref{fig:transition_new_favorites} shows that the introduction of \TEC leads to a rapid \emph{decline} in new favorites in the transition simulation, which stabilizes at a level substantially below the steady state under \Greedy after about 25 rounds. Since \Greedy is designed to maximize the incidence of favoriting, whereas \TEC prioritizes templates with greater labor demand, a reduction in the number of new favorites is a natural prediction of the model. From this perspective, the absence of such a decline in the field experiment can be viewed as an unexpected success.

As suggested by the impression-level analysis (Table~\ref{tb:impression_did_combined}), this discrepancy is likely driven by the fact that \TEC reduces repeat views of the same template and increases the favoriting probability per impression, offsetting the mechanical reduction in exposure to highly favorited templates. By contrast, the simulation assumes that the probability of favoriting upon viewing a template is fixed, thereby abstracting from behavioral responses such as changes in repeat exposure or per-impression engagement.

\paragraph{Matches}
Second, we turn to matches between workers and offerings. In the field experiment, a statistically significant increase in realized matches was already observed within one month. Figure~\ref{fig:transition_worker_activation} shows that this effect should continue to build under a longer rollout. In the model, the cumulative mean of per-round job-finding rate under \TEC begins to exceed that under \Greedy almost immediately after the switch but approaches its long-run level only gradually. This gradual adjustment arises because, while the set of templates newly favorited by workers responds quickly to the change in recommendations, the overall composition of workers' favorite lists evolves more slowly. As a result, matching continues to improve until favorite lists reach their steady-state distribution. Consistent with this mechanism, roughly 60\% of the eventual cumulative-mean gain is reflected by around round 25 and roughly 85\% by around round 75. The one-month rollout is therefore best interpreted as capturing only the early or middle part of the adjustment path for matching-related outcomes. If \TEC had remained in place longer, the model predicts larger gains in realized matching than those observed during the field experiment.

\paragraph{Recommendations per Active Template}
We now focus on the effect of \TEC in concentrating exposure on active templates. We begin by examining how recommendations are allocated across templates with different levels of activity. In the field results, panel (b) of Table~\ref{tb:prefecture_day_did_combined} reports statistics computed over templates that were recommended at least once on a given day, due to data limitations. In the transition simulation, we adopt a more direct measure of template activity and define a template as active if it posts, on average, at least 0.1 offerings per round.

In the field experiment, even within a one-month rollout, we observe a clear shift in recommendations toward active templates. The transition simulation results shown in Figure~\ref{fig:transition_recommendations_per_active_template} are consistent with this pattern: following the introduction of \TEC, exposure rapidly shifts toward active templates because this is the most direct empirical signature of the algorithm's exposure-reallocation mechanism. This effect largely saturates by around round 25, converging to the steady-state level under \TEC. This suggests that, under a longer deployment, \TEC\ would sustain a similar degree of concentration of recommendations on active templates as observed in the field experiment.

\paragraph{Fulfillment per Active Template}
Finally, we analyze the effect of \TEC in increasing the likelihood that offerings from active templates are successfully filled. The field rollout did not produce a detectable increase in active-offering fulfillment. The transition simulation shown as Figure~\ref{fig:transition_fulfillment_per_active_template} helps interpret this null as well. Because exposure allocation changes first, and subscribers and per-round fill rate accumulate only gradually, concentration of fulfillment among active offerings is also expected to emerge with a delay. This timing argument is especially relevant because baseline fulfillment in the experimental prefectures was already high, leaving limited room for short-run gains. The steady-state simulation nevertheless predicts sizable fulfillment improvements when \TEC is operated long enough, especially in environments where baseline fulfillment under \Greedy is lower. A longer rollout would therefore be more likely to reveal fulfillment gains, particularly in markets with more slack than those observed in Aomori and Iwate.

\bigskip

As an empirical anchor, the per-round viewing, search, replacement, and offering-arrival probabilities are calibrated to daily platform frequencies. Under this calibration, the simulation suggests that recommendation flows should respond quickly, while subscriber stocks and fulfillment evolve over a longer horizon and do not saturate even after 200 rounds. Nevertheless, precise mapping between calendar time in the field experiment and rounds in the simulation is difficult to establish. In the simulation, worker replacement, recommendation of favorite candidates, favoriting, and matching all occur within each round, whereas in the field setting, these processes unfold on different time scales. 

Despite this mismatch, the qualitative dynamics are informative for favorite-list management. Effects on new favorites and recommendation allocation appear immediately and strongly, while match-related outcomes emerge more gradually, as they depend on the evolution of workers' favorite lists. Taken together, the combined evidence suggests that the live rollout captured only the early phase of the transition to \TEC rather than its full market-level impact. In other words, the benefits of \TEC are expected to become more pronounced over time, not because the algorithm continues to alter exposure in new ways, but because the consequences of its initial reallocation propagate through favorite formation, subscriber stocks, and subsequent matching opportunities.

\section{Conclusion}

This paper studies recommender system design in spot-work markets through the lens of exposure allocation. Recommending templates that users are most likely to favorite does not necessarily improve realized matching outcomes. Using institutional features of Timee as motivation, we show how a conventional recommender, \Greedy, can induce misdirected concentration of exposure, directing recommendations toward templates that attract attention but generate limited labor demand. This misallocation can harm both sides of the market by reducing workers' effective access to viable opportunities and increasing the risk of unfilled offerings for templates with substantial demand.

To mitigate these inefficiencies, we develop quota-based recommender systems that control exposure at the template level. \SQ illustrates the value of limiting concentration, but its fixed quota is difficult to tune across market environments. We therefore propose \AQ, which directs recommendations toward templates with greater unmet labor demand using recent posting activity and unfilled capacity. Our main deployable algorithm, \TEC, implements this idea through a fully parallelizable threshold rule suitable for large-scale, real-time recommendation. In calibrated simulations using detailed operational data from Timee, \TEC raises the per-round job-finding rate from 57.6\% under \Greedy to 70.0\%, closely matching the sequential \AQ benchmark and robustly outperforming \Greedy and \SQ across market conditions.

The field experiment provides real-world evidence consistent with this logic and clarifies the short-run transition induced by \TEC. In the randomized prefecture rollout, \TEC increases realized matches at the prefecture-day level and raises the average number of recommended workers per active template. A distributional analysis shows that this reallocation appears mainly as a reduction in the lower tail of recommendation counts across active template-days. For subscribers per template, the same analysis shows a lower zero-subscriber mass, even though average subscriber stocks and active-offering fulfillment do not shift substantially within the short rollout window. This pattern is consistent with the mechanism: these outcomes depend on the gradual accumulation and use of favorite lists rather than on contemporaneous exposure alone. Transition simulations indicate that a longer deployment would allow the exposure reallocation to propagate more fully into subscriber accumulation and downstream fulfillment. The live deployment also adds evidence that the simulation cannot provide: at the impression level, treated recommendation exposures become more likely to generate favorites and downstream matches, suggesting that \TEC improved the quality of the recommendation lists presented to users in addition to changing aggregate exposure allocation.

Several limitations point to directions for future work. First, our analysis emphasizes worker-to-offering matches as a primary market-level outcome. While this metric is natural and important for spot-work markets, it does not capture all dimensions of welfare, such as heterogeneity in job quality, worker surplus, or distributional impacts across worker types and firms. Second, our analysis abstracts from some channels through which workers discover offerings, including browsing and other interface surfaces, and from potential longer-run responses by workers and firms. Extending the framework to incorporate multi-channel discovery, worker learning, and endogenous firm posting decisions would further clarify the welfare effects of exposure allocation policies. Third, our field experiment studies a short rollout of \TEC. As our transition simulation suggests, a longer deployment may generate larger effects on subscriber accumulation, fulfillment, and downstream matching outcomes than those observed during the experimental window. Estimating these long-run effects in a fully controlled field experiment remains an important question, although such a design is difficult to implement on a live platform because experimental control must be balanced against operational and business constraints. Finally, the broader lesson of the paper extends beyond spot work: in many digitally mediated markets, recommender systems operate as allocation mechanisms under capacity constraints, and designing them to optimize market-level outcomes, rather than myopic conversion, can yield substantial welfare gains.

\bibliographystyle{aer}
\bibliography{utmd_timee}

\newpage
\appendix
\section{Additional Figures and Tables}

\subsection{Position Effects in Impression-Level Analysis}

\begin{figure}[!tp]
\centering
\includegraphics[width=0.85\textwidth]{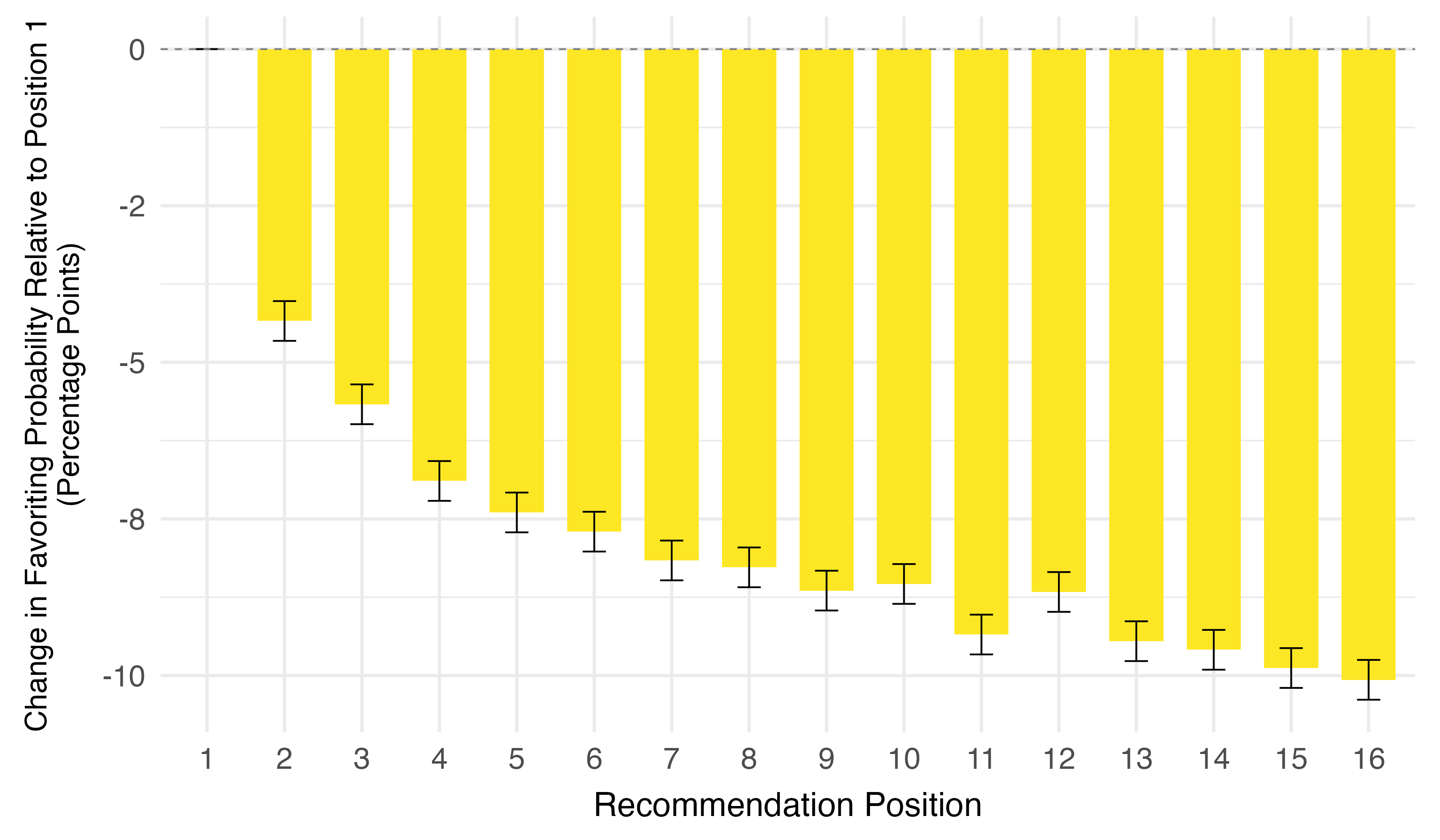}
\caption{Estimated Recommendation-Position Fixed Effects on Favoriting Probability}
\label{fig:position_fe_impression_favorite}
\end{figure}

Figure~\ref{fig:position_fe_impression_favorite} reports the estimated position fixed effects from the impression-level favorite regression. Each bar shows the association between recommendation position $k$ and favoriting probability relative to position 1 (top of the list), with 95\% confidence intervals. The favorite probability falls by approximately 10 percentage points from position 1 to position 16, with small nonmonotonic differences across adjacent ranks. Because within-list ranking was not randomized, these estimates combine possible display-order effects with systematic differences in the templates shown at different ranks. They should therefore be read as descriptive evidence consistent with position bias, not as causal position effects. Our experiment varies the \emph{composition} of the recommendation set, but not the ranking within it; optimizing the within-list ordering is a complementary design lever that could further improve matching outcomes.

\subsection{Distributional Results}

\begin{figure}[!tp]
  \begin{center}
      \begin{subfigure}{0.8\textwidth}
        \centering
        \includegraphics[width=\textwidth]{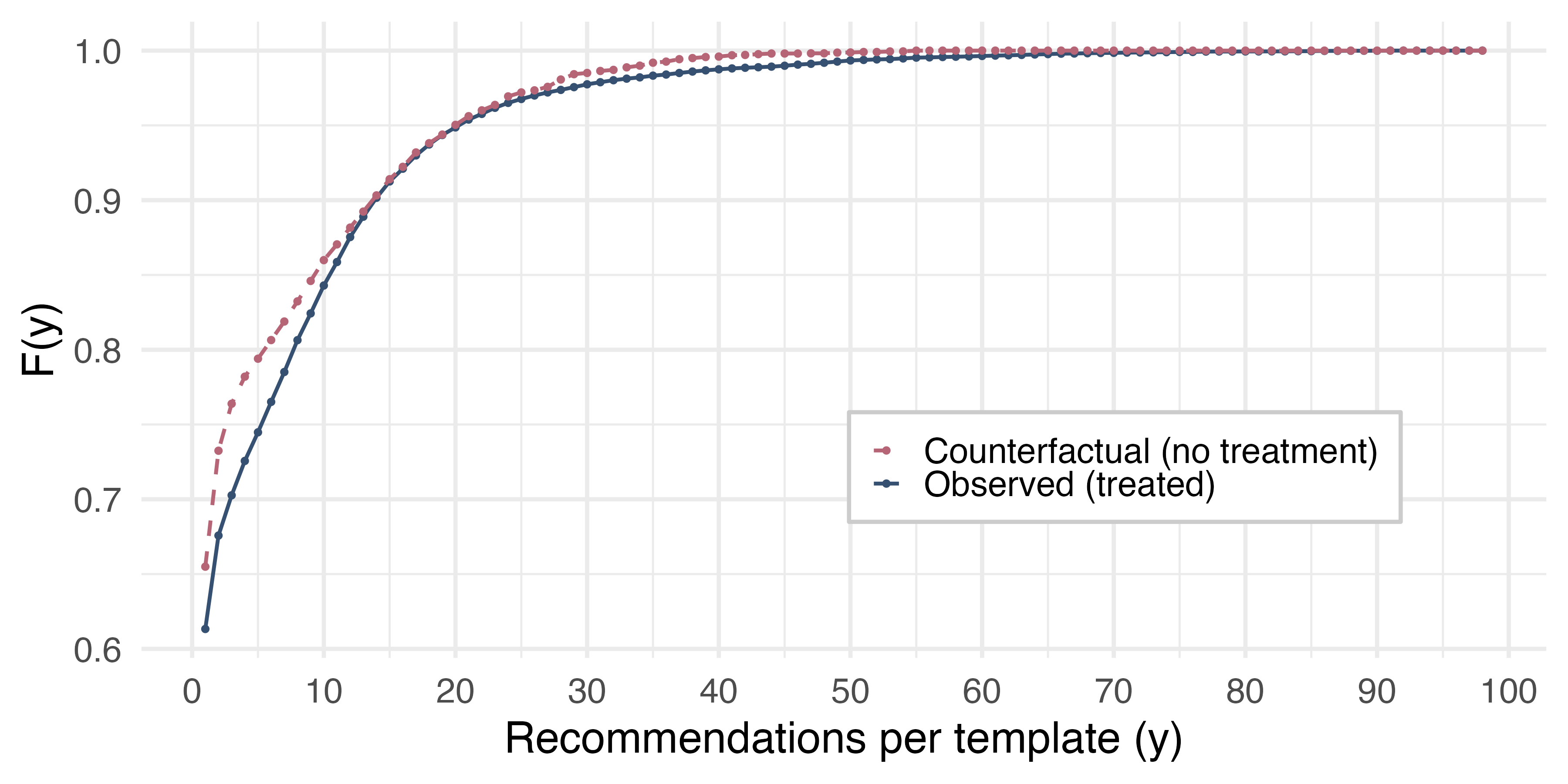}
        \caption{Recommendations per Template}
      \end{subfigure}\\
      \bigskip
      \begin{subfigure}{0.8\textwidth}
        \centering
        \includegraphics[width=\textwidth]{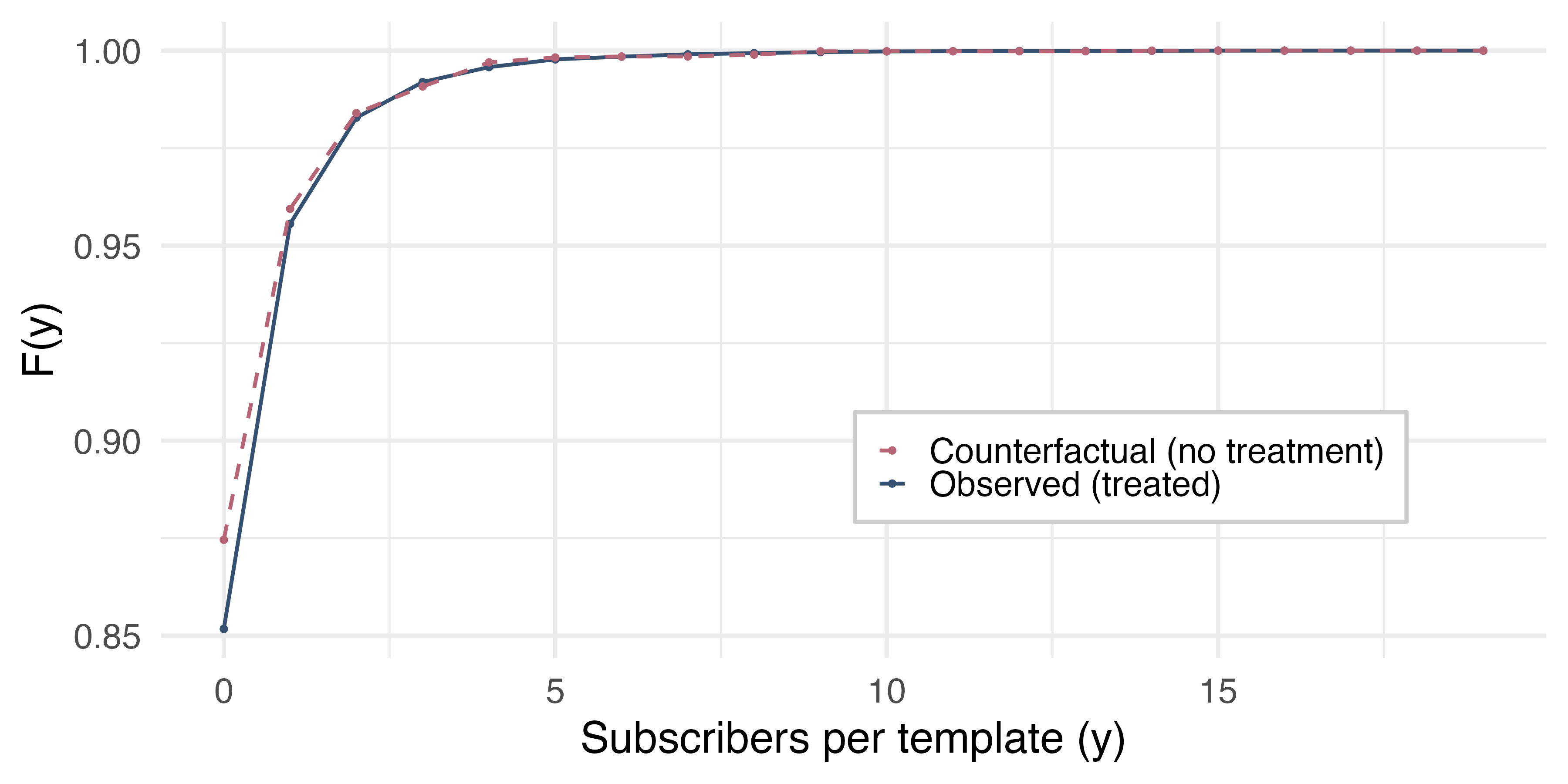}
        \caption{Subscribers per Template}
      \end{subfigure}
  \end{center}
  \caption{Observed vs.\ Counterfactual CDFs of Template-Day-Level Outcomes}\footnotesize
  \textit{Note}: Solid lines plot the observed CDF $\hat{F}_{Y_1|G,T}(y|1,1)$ of treated templates in the post-period. Dashed lines plot the counterfactual CDF $\hat{F}_{Y_0|G,T}(y|1,1)$ recovered from the distribution-regression DID. CDFs are monotonized following Algorithm~1 of \cite{fernandez2024distribution}.
  \label{fig:drdid_cdf_actual_counterfactual}
\end{figure}

Figure~\ref{fig:drdid_cdf_actual_counterfactual} plots the underlying CDFs from which the treatment effects in Figure~\ref{fig:drdid_delta_f} are derived. Panel~(a) confirms that the observed CDF of recommendations per template under \TEC lies strictly below the counterfactual throughout the lower portion of the support: treated templates are less likely to fall into low-recommendation bins than they would have been absent treatment. The gap is largest for thresholds between 3 and 10, consistent with the peak of $\widehat{\Delta}_F(y)$ in Figure~\ref{fig:drdid_delta_f}(a). Panel~(b) shows that the two CDFs for subscribers per template are nearly indistinguishable, reinforcing the null distributional result. The figure also suggests a possible reason: in this rural area over a short intervention period, subscriber counts are heavily concentrated at 1, 2, and 3, so that the CDF rises steeply to near unity by $y=5$, potentially leaving too little distributional variation for any treatment effect to be detected.

\section{Recommender Details}\label{sec:recommender_details}

\subsection{Score Capping for \TEC}\label{subsec:score_capping}

For \TEC, we apply \emph{score capping} as a pre-processing step before computing eligibility thresholds from scores. It caps scores to prevent the choice set from becoming empty because a single template holds an excessively large score. Specifically, the procedure prevents a high-scoring template from receiving an eligibility-threshold interval $\tau_{\pi(j)} - \tau_{\pi(j+1)}$ that exceeds $1/c$. Assigning an interval longer than $1/c$ does not create an additional useful opportunity because the same template cannot be recommended twice to the same worker in one recommendation list. Instead, the extra interval length only increases the chance that the choice set becomes empty. For example, if a worker reaches another selection timing inside an interval that points to a template already included earlier in the same recommendation list, that template is no longer feasible. In that case, the threshold rule may have no feasible candidate left and must rely on the greedy completion rule. To avoid this, \TEC caps scores so that every template's marginal eligibility interval has length at most $1/c$.

Define the score cap by $p^{\max} = P/c$, where $P$ denotes the total score. We construct \emph{capped scores} $\bar p_k$ such that no template has a score greater than $p^{\max}$, while preserving the relative balance of scores across the other templates. The capping procedure does not discard the excess score. Instead, whenever a template's score exceeds $p^{\max}$, the excess score is redistributed to templates that have not yet reached the cap. The redistribution is proportional to their current scores. This rule follows the interpretation of scores as exposure demand: among templates that are still below the cap, a template with twice the current score receives twice as much of the redistributed excess. Thus, the procedure preserves the relative priority implied by the original scores as much as possible, while preventing any single template from taking more than one slot's worth of eligibility mass.

Set $p^{(0)}_k=p_k$ for all $k \in K$. We scan templates in the order $\pi(1),\ldots,\pi(|K|)$. At step $j$, if $p^{(j-1)}_{\pi(j)} \le p^{\max}$, then the procedure stops. Since the proportional redistribution preserves the relative order of the remaining templates, all templates not yet scanned also have scores weakly below $p^{\max}$.

If instead $p^{(j-1)}_{\pi(j)} > p^{\max}$, template $\pi(j)$ is capped at $p^{\max}$. The excess score is $e^{(j)} = p^{(j-1)}_{\pi(j)} - p^{\max}$. Let $S^{(j)} = \sum_{h=j+1}^{|K|} p^{(j-1)}_{\pi(h)}$ be the total current score of templates that have not yet been scanned. When $S^{(j)}>0$, the score vector is updated as follows:
\begin{equation}
    p^{(j)}_{\pi(h)}
    =
    \begin{cases}
    p^{(j-1)}_{\pi(h)},
        & h<j,\\[4pt]
    p^{\max},
        & h=j,\\[4pt]
    p^{(j-1)}_{\pi(h)}
    \left(
        1+\dfrac{e^{(j)}}{S^{(j)}}
    \right),
        & h>j.
    \end{cases}
\end{equation}
Thus, all remaining templates are multiplied by the same factor. This keeps their relative scores unchanged. The algorithm then proceeds to step $j+1$.

Let $\bar p_k$ denote the capped score obtained when the procedure stops. By construction, the capped scores preserve the total score: $\sum_{k \in K} \bar p_k = \sum_{k \in K} p_k = P$. They also satisfy the cap: $\bar{p}_k \le P/c$ for all $k \in K$. The capped scores are used only to construct eligibility thresholds in the current round. The underlying score state $p_k$ is updated according to the score update rule.

\begin{algorithm}[h!] \caption{\TEC: Score Capping} \label{alg:score_capping} 
\begin{algorithmic}[1] 
\Statex \textbf{Input:} Raw scores $(p_k)_{k\in K}$, recommendation capacity $c$ 
\Statex \textbf{Output:} Capped scores $(\bar p_k)_{k\in K}$ 
\Statex 
\For{each template $k\in K$} 
    \State $\bar p_k \leftarrow \max\{p_k,0\}$ 
\EndFor 
\State $\tilde P \leftarrow \sum_{k\in K}\bar p_k$ 
\If{$\tilde P=0$} 
    \State \textbf{return} $(\bar p_k)_{k\in K}$ 
\EndIf 
\State $p^{\max}\leftarrow \tilde P/c$ 
\State $A\leftarrow K$ \Comment{Templates not yet fixed at the cap}
\While{there exists $k\in A$ such that $\bar p_k>p^{\max}$} 
    \State Choose $k^\ast\in \argmax_{k\in A}\bar p_k$ 
    \State $e\leftarrow \bar p_{k^\ast}-p^{\max}$ \Comment{Excess score} 
    \State $\bar p_{k^\ast}\leftarrow p^{\max}$ 
    \State $A\leftarrow A\setminus\{k^\ast\}$ 
    \State $\bar{p}_A\leftarrow \sum_{l\in A}\bar p_l$ 
    \For{each $l \in A$} 
    \State $\bar p_l \leftarrow \bar p_l + e\cdot \bar p_l/\bar{p}_A$ 
    \EndFor 
\EndWhile 
\State \textbf{return} $(\bar p_k)_{k\in K}$ 
\end{algorithmic} 
\end{algorithm}

\subsection{Pseudocode}\label{subsec:pseudo_code}

This section provides the pseudocode for the recommenders analyzed in the main text.

The set of workers is denoted by $I$, and the set of templates is denoted by $K$. Each worker $i \in I$ has a vector of predicted favoriting probabilities $(\alpha_{\theta(i)\theta(k)})_{k\in K}$. For each worker $i$, let $F_i \subseteq K$ be the set of templates already in their favorite list, and $K_i \subseteq K \setminus F_i$ be the feasible set of templates that are eligible to be recommended. Let $c$ be the recommendation capacity, which is the maximum number of templates to be recommended to each worker. The goal of each algorithm is to output a recommendation set $R_i$ for each worker $i$, where $|R_i| \leq c$.


\subsubsection{Greedy}
\Greedy serves as our baseline. It simply sorts the feasible templates for each worker in descending order of their predicted favoriting probabilities and recommends the top $c$ templates.

\begin{algorithm}[h!]
\caption{Greedy}
\label{alg:greedy}
\begin{algorithmic}[1]
\Statex \textbf{Input:} Predicted favoriting probabilities $(\alpha_{\theta(i)\theta(k)})_{i \in I, k \in K}$, 
\Statex \hspace{\algorithmicindent} Feasible template sets $(K_i)_{i \in I}$, 
\Statex \hspace{\algorithmicindent} Recommendation capacity $c$
\Statex \textbf{Output:} Recommendation lists for all workers $(R_i)_{i \in I}$
\Statex

\For{each worker $i \in I$}
    \State $R_i \leftarrow \emptyset$
    \State Sort $K_i$ in descending order of $\alpha_{\theta(i)\theta(k)}$
    \For{each $k \in K_i$}
        \If{$|R_i| = c$}
            \State \textbf{break}
        \EndIf
        \State $R_i \leftarrow R_i \cup \{k\}$
    \EndFor
\EndFor
\State \textbf{return} $(R_i)_{i \in I}$
\end{algorithmic}
\end{algorithm}


\subsubsection{Round-Robin RSD}

The following algorithm is the common allocation subroutine used by \SQ and \AQ.
Each template $k$ is assigned an integer quota $q_k$, interpreted as the maximum number of recommendation exposures that template $k$ can receive in the current construction. 
Workers are randomly ordered once, and the same order is used in every slot $s=1,\ldots,c$.

\begin{algorithm}[h!]
\caption{Round-Robin RSD with Template Quotas}
\label{alg:round_robin_rsd}
\begin{algorithmic}[1]
\Statex \textbf{Input:} Predicted favoriting probabilities $(\alpha_{\theta(i)\theta(k)})_{i\in I,k\in K}$,
\Statex \hspace{\algorithmicindent} Feasible template sets $(K_i)_{i\in I}$, 
\Statex \hspace{\algorithmicindent} Recommendation capacity $c$,
\Statex \hspace{\algorithmicindent} Template quotas $(q_k)_{k\in K}$, where $q_k\in\mathbb{Z}_{+}\cup\{+\infty\}$
\Statex \textbf{Output:} Recommendation lists $(R_i)_{i\in I}$
\Statex

\For{each template $k\in K$}
    \State $n_k\leftarrow 0$ \Comment{Realized recommendation count}
\EndFor

\For{each worker $i\in I$}
    \State $R_i\leftarrow \emptyset$
    \State Sort $K_i$ in descending order of $\alpha_{\theta(i)\theta(k)}$
\EndFor

\State Draw a uniformly random permutation $(i_1,\ldots,i_{|I|})$ of workers

\For{$r=1,\ldots,c$}
    \For{each worker $i$ in the order of $(i_1,\ldots,i_{|I|})$}
        \State Find the first template $k\in K_i$ such that $k\notin R_i$ and $n_k < q_k$
        \If{such $k$ exists}
            \State $R_i\leftarrow R_i\cup\{k\}$
            \State $n_k\leftarrow n_k+1$
        \EndIf
    \EndFor
\EndFor

\State \textbf{return} $(R_i)_{i\in I}$
\end{algorithmic}
\end{algorithm}

\subsubsection{Static Quota (SQ)}
\SQ sets a common quota $q$ for every template and then runs round-robin RSD.

\begin{algorithm}[h!]
\caption{Static Quota (\SQ)}
\label{alg:sq}
\begin{algorithmic}[1]
\Statex \textbf{Input:} Predicted favoriting probabilities $(\alpha_{\theta(i)\theta(k)})_{i\in I,k\in K}$,
\Statex \hspace{\algorithmicindent} Feasible template sets $(K_i)_{i\in I}$
\Statex \hspace{\algorithmicindent} Recommendation capacity $c$,
\Statex \hspace{\algorithmicindent} Common quota $q$
\Statex \textbf{Output:} Recommendation lists $(R_i)_{i\in I}$
\Statex

\For{each template $k\in K$}
    \State $q_k\leftarrow q$
\EndFor

\State $(R_i)_{i\in I}
    \leftarrow
    \Call{Round-Robin RSD}{(\alpha_{\theta(i)\theta(k)}),(K_i),c,(q_k)}$

\State \textbf{return} $(R_i)_{i\in I}$
\end{algorithmic}
\end{algorithm}

\subsubsection{Adaptive Quota (AQ)}
\AQ computes template-specific quotas from recent posting activity and unfilled capacity, and then runs round-robin RSD. 
The variables $x_k^{0,(t-1)}$ and $x_k^{1,(t-1)}$ denote, respectively, the posted offering capacity and unfilled capacity of template $k$ observed in the previous round. 
Thus, the quota computed below is used to generate recommendations in the current round $t$.

\begin{algorithm}[h!]
\caption{Adaptive Quota (\AQ)}
\label{alg:aq}
\begin{algorithmic}[1]
\Statex \textbf{Input:} Predicted favoriting probabilities $(\alpha_{\theta(i)\theta(k)})_{i\in I,k\in K}$,
\Statex \hspace{\algorithmicindent} Feasible template sets $(K_i)_{i\in I}$,
\Statex \hspace{\algorithmicindent} Recommendation capacity $c$,
\Statex \hspace{\algorithmicindent} Average quota $\bar q$,
\Statex \hspace{\algorithmicindent} Score weights $w^0,w^1$,
\Statex \hspace{\algorithmicindent} Previous-round posted capacity $(x_k^{0,(t-1)})_{k\in K}$,
\Statex \hspace{\algorithmicindent} Previous-round unfilled capacity $(x_k^{1,(t-1)})_{k\in K}$
\Statex \textbf{Output:} Recommendation lists $(R_i)_{i\in I}$
\Statex

\For{each template $k\in K$}
    \State $p_k\leftarrow w^0 x_k^{0,(t-1)} + w^1 x_k^{1,(t-1)}$
\EndFor

\State $p_K\leftarrow \sum_{k\in K}p_k$

\If{$p_K=0$}
    \For{each template $k\in K$}
        \State $q_k \leftarrow \lfloor \bar{q} \rfloor$
    \EndFor
\Else
    \For{each template $k\in K$}
        \State $q_k\leftarrow \lfloor \bar q |K|\cdot p_k/p_K \rfloor$
    \EndFor
\EndIf
\State $(R_i)_{i\in I}
    \leftarrow
    \Call{Round-Robin RSD}{(\alpha_{\theta(i)\theta(k)}),(K_i),c,(q_k)}$

\State \textbf{return} $(R_i)_{i\in I}$
\end{algorithmic}
\end{algorithm}

\subsubsection{Thresholded Eligibility Control (TEC)}

\TEC comprises three parts: threshold construction, recommendation generation, and score update. The threshold construction part converts each template's score into an eligibility threshold. The recommendation generation phase then computes the recommendation lists based on workers' predicted favoriting probabilities and the thresholds. After the matching phase, the score update phase computes the next-round scores based on the observed posted capacity and unfilled capacity.

In the recommendation generation part, \TEC constructs the recommendation list $R_i$ for each worker independently and in parallel. To approximate the sequential nature of RSD, the system first draws a random permutation of workers to determine each worker's selection timing $z_{is}$ across slots. Crucially, because this selection timing is monotonically increasing with respect to the slot index $s$, any template that fails the eligibility condition ($\tau_k \ge z_{is}$) for a given slot will inevitably fail for all subsequent slots. 
Conditional on the worker-specific candidate list being sorted by predicted favoriting probability,
the thresholded selection step requires only a single pass and runs in $O(|K_i|)$ time per worker.
Any remaining unfilled slots are handled by a greedy fallback completion rule.

\begin{algorithm}[h!]
\caption{Thresholded Eligibility Control (\TEC)}
\label{alg:tec_one_round}
\begin{algorithmic}[1]
\Statex \textbf{Input:} Current raw scores $(p_k^{(t)})_{k\in K}$,
\Statex \hspace{\algorithmicindent} Predicted favoriting probabilities $(\alpha_{\theta(i)\theta(k)})_{i\in I,k\in K}$,
\Statex \hspace{\algorithmicindent} Feasible template sets $(K_i)_{i\in I}$,
\Statex \hspace{\algorithmicindent} Recommendation capacity $c$,
\Statex \hspace{\algorithmicindent} Score weights $w^0,w^1$
\Statex \textbf{Output:} Recommendation lists $(R_i^{(t)})_{i\in I}$,
\Statex \hspace{\algorithmicindent} Next-round scores $(p_k^{(t+1)})_{k\in K}$
\Statex

\Statex \textbf{Threshold construction}
\State $(\tau_k^{(t)})_{k\in K}
    \leftarrow
    \Call{TEC: Threshold Construction}{(p_k^{(t)}),c}$

\Statex \textbf{Recommendation generation}
\State $(R_i^{(t)})_{i\in I}
    \leftarrow
    \Call{TEC: Recommendation Generation}{(\tau_k^{(t)}),(\alpha_{\theta(i)\theta(k)}),(K_i),c}$

\State Deliver recommendations $(R_i^{(t)})_{i\in I}$
\State After the matching phase, observe posted capacity $(x_k^{0,(t)})_{k\in K}$ and unfilled capacity $(x_k^{1,(t)})_{k\in K}$

\Statex \textbf{Score update}
\For{each template $k\in K$} \Comment{Recommendation counts}
    \State $r_k^{(t)}\leftarrow \sum_{i\in I}\mathbf{1}\{k\in R_i^{(t)}\}$
\EndFor
\State $(p_k^{(t+1)})_{k\in K}
    \leftarrow
    \Call{TEC: Score Update}{(p_k^{(t)}),(r_k^{(t)}),(x_k^{0,(t)}),(x_k^{1,(t)}),w^0,w^1}$

\State \textbf{return} $(R_i^{(t)})_{i\in I},(p_k^{(t+1)})_{k\in K}$
\end{algorithmic}
\end{algorithm}

\begin{algorithm}[h!]
\caption{\TEC: Threshold Construction}
\label{alg:tec_thresholds}
\begin{algorithmic}[1]
\Statex \textbf{Input:} Current raw scores $(p_k)_{k\in K}$, 
\Statex \hspace{\algorithmicindent} Recommendation capacity $c$
\Statex \textbf{Output:} Eligibility thresholds $(\tau_k)_{k\in K}$
\Statex

\State $(\bar p_k)_{k\in K}\leftarrow \Call{TEC: Score Capping}{(p_k),c}$
\State $\bar{p}_K \leftarrow \sum_{k\in K}\bar p_k$


\State Sort templates so that $\bar p_{\pi(1)}\ge \bar p_{\pi(2)}\ge \cdots \ge \bar p_{\pi(|K|)}$

\For{$j=1,\ldots,|K|$}
    \State $\displaystyle
        \tau_{\pi(j)}
        \leftarrow
        \frac{1}{\bar{p}_K}\sum_{h=j}^{|K|}\bar p_{\pi(h)}
    $
\EndFor

\State \textbf{return} $(\tau_k)_{k\in K}$
\end{algorithmic}
\end{algorithm}

\begin{algorithm}[h!]
\caption{\TEC: Recommendation Generation}
\label{alg:tec_recommend}
\begin{algorithmic}[1]
\Statex \textbf{Input:} Eligibility thresholds $(\tau_k)_{k\in K}$,
\Statex \hspace{\algorithmicindent} Predicted favoriting probabilities $(\alpha_{\theta(i)\theta(k)})_{i\in I,k\in K}$,
\Statex \hspace{\algorithmicindent} Feasible template sets $(K_i)_{i\in I}$,
\Statex \hspace{\algorithmicindent} Recommendation capacity $c$
\Statex \textbf{Output:} Recommendation lists $(R_i)_{i\in I}$
\Statex

\State Draw a uniformly random bijection $\rho:I\to\{1,\ldots,|I|\}$

\For{each worker $i\in I$ \textbf{in parallel}}
    \State $R_i\leftarrow \emptyset$
    \State $s\leftarrow 1$
    \State Sort $K_i$ in descending order of $\alpha_{\theta(i)\theta(k)}$.
    \For{each template $k\in K_i$}
        \If{$s>c$}
            \State \textbf{break}
        \EndIf

        \State $\displaystyle
            z_{is}\leftarrow
            \frac{1}{c}
            \left(
                s-1+\frac{\rho(i)-1}{|I|}
            \right)
        $

        \If{$\tau_k\ge z_{is}$}
            \State $R_i\leftarrow R_i\cup\{k\}$
            \State $s\leftarrow s+1$
        \EndIf
    \EndFor

    \For{each template $k\in K_i$} \Comment{Greedy fallback completion}
        \If{$|R_i|=c$}
            \State \textbf{break}
        \EndIf
        \If{$k\notin R_i$}
            \State $R_i\leftarrow R_i\cup\{k\}$
        \EndIf
    \EndFor
\EndFor

\State \textbf{return} $(R_i)_{i\in I}$
\end{algorithmic}
\end{algorithm}

\begin{algorithm}[h!]
\caption{\TEC: Score Update}
\label{alg:tec_update_scores}
\begin{algorithmic}[1]
\Statex \textbf{Input:} Current raw scores $(p_k^{(t)})_{k\in K}$,
\Statex \hspace{\algorithmicindent} Recommendation counts $(r_k^{(t)})_{k\in K}$,
\Statex \hspace{\algorithmicindent} Posted capacity $(x_k^{0,(t)})_{k\in K}$,
\Statex \hspace{\algorithmicindent} Unfilled capacity $(x_k^{1,(t)})_{k\in K}$,
\Statex \hspace{\algorithmicindent} Score weights $w^0,w^1$
\Statex \textbf{Output:} Next-round raw scores $(p_k^{(t+1)})_{k\in K}$
\Statex

\For{each template $k\in K$}
    \State $p_k^{+,(t)}\leftarrow w^0 x_k^{0,(t)}+w^1 x_k^{1,(t)}$
\EndFor

\State $p_K^{+,(t)}\leftarrow \sum_{k\in K}p_k^{+,(t)}$
\State $r_K^{(t)} \leftarrow \sum_{k \in K}r_k^{(t)}$

\For{each template $k\in K$}
    \State $\displaystyle
        p_k^{-,(t)}
        \leftarrow
        \frac{r_k^{(t)}}{r_K^{(t)}}p_K^{+,(t)}
    $
    \State $\displaystyle
        p_k^{(t+1)}
        \leftarrow
        p_k^{(t)}+p_k^{+,(t)}-p_k^{-,(t)}
    $
\EndFor

\State \textbf{return} $(p_k^{(t+1)})_{k\in K}$
\end{algorithmic}
\end{algorithm}

\end{document}